\documentclass[chicago, iop, apj]{emulateapj}

\usepackage{natbib}
\usepackage{bm}
\usepackage{times}
\usepackage{graphicx}
\usepackage{color}
\usepackage{amsmath}
\usepackage{amssymb}
\usepackage{fancyvrb}
\usepackage{url}
\usepackage{hyperref}  
\usepackage{verbatim}

\newcommand{\mat}[1]{{\bf {#1}}}
\newcommand{\vect}{\boldsymbol}

\renewcommand{\cite}{\citep}

\newcommand{\CII}{C$\,${\sc ii}}

\newcommand{\fgfmodes}{\mat{U}_{\rm f}}
\newcommand{\fgsmodes}{\mat{V}_{\rm f}}
\newcommand{\fgmap}{\mat{X}_{\rm f}}

\newcommand{\sfmodes}{\mat{U}_{\rm s}}

\newcommand{\smap}{\mat{X}_{\rm s}}

\newcommand{\nmap}{\mat{X}_{\rm n}}

\newcommand{\qed}{\nobreak \ifvmode \relax \else
      \ifdim\lastskip<1.5em \hskip-\lastskip
      \hskip1.5em plus0em minus0.5em \fi \nobreak
      \vrule height0.75em width0.5em depth0.25em\fi}

\begin{document}
\VerbatimFootnotes

\slugcomment{Submitted to ApJ}

\author{
E.~R.~Switzer\altaffilmark{1}, T.-C.~Chang\altaffilmark{2}, K.~W.~Masui\altaffilmark{3,4}, U.-L.~Pen\altaffilmark{5,4}, T.~C.~Voytek\altaffilmark{6, 7}
}

\altaffiltext{1}{NASA Goddard Space Flight Center, Greenbelt, MD 20771, USA}
\email{eric.r.switzer@nasa.gov}
\altaffiltext{2}{Academia Sinica Institute of Astronomy and Astrophysics 
11F of Astro-Math Building, AS/NTU, 1 Roosevelt Rd Sec 4, Taipei, 10617, Taiwan}
\altaffiltext{3}{Department of Physics and Astronomy, University of British Columbia, 6224 Agricultural Rd. Vancouver, V6T 1Z1, Canada}
\altaffiltext{4}{Canadian Institute for Advanced Research, CIFAR Program in Cosmology and Gravity, Toronto, ON, M5G 1Z8}
\altaffiltext{5}{Canadian Institute for Theoretical Astrophysics, University of Toronto, 60 St. George St., Toronto, ON M5S 3H8, Canada}
\altaffiltext{6}{McWilliams Center for Cosmology, Carnegie Mellon University, Department of Physics, 5000 Forbes Ave., Pittsburgh, PA 15213, USA}
\altaffiltext{7}{Astrophysics and Cosmology Research Unit, School of Chemistry and Physics, University of KwaZulu-Natal, Durban, 4041, South Africa}

\shortauthors{Switzer}

\title{INTERPRETING THE UNRESOLVED INTENSITY OF COSMOLOGICALLY REDSHIFTED LINE RADIATION
\shorttitle{Methods for single-dish intensity mapping}}

\begin{abstract}
Intensity mapping experiments survey the spectrum of diffuse line radiation rather than detect individual objects at high signal-to-noise ratio. Spectral maps of unresolved atomic and molecular line radiation contain three-dimensional information about the density and environments of emitting gas and efficiently probe cosmological volumes out to high redshift. Intensity mapping survey volumes also contain all other sources of radiation at the frequencies of interest. Continuum foregrounds are typically $\sim 10^2$--$10^3$ times brighter than the cosmological signal.  The instrumental response to bright foregrounds will produce new spectral degrees of freedom that are not known in advance, nor necessarily spectrally smooth. The intrinsic spectra of foregrounds may also not be well known in advance. We describe a general class of quadratic estimators to analyze data from single-dish intensity mapping experiments and determine contaminated spectral modes from the data themselves. The key attribute of foregrounds is not that they are spectrally smooth, but instead that they have fewer bright spectral degrees of freedom than the cosmological signal. Spurious correlations between the signal and foregrounds produce additional bias. Compensation for signal attenuation must estimate and correct this bias. A successful intensity mapping experiment will control instrumental systematics that spread variance into new modes, and it must observe a large enough volume that contaminant modes can be determined independently from the signal on scales of interest.
\end{abstract}

\maketitle

\section{Introduction}

Intensity mapping is an emerging technique for cosmological observation. It uses atomic or molecular transition radiation to tomographically map large volumes of the universe. These data volumes contain a combination of information about the abundance, environment, and velocity of emitters. Tomographic line surveys offer the potential to capture the dynamic universe in epochs that are otherwise difficult to observe \citep{1997ApJ...475..429M} and to probe a variety of galactic environments in aggregate \citep{2014ApJ...793..116U, 2015arXiv150308833L}. The recovery of many modes could allow intensity surveys to compete with dark energy constraints from standard spectroscopic galaxy surveys \citep{2008PhRvL.100i1303C, 2008PhRvL.100p1301L}. 

Intensity mapping shares some parallels with studies of cosmic background radiation and spectroscopic galaxy surveys. Like background radiation studies, intensity mapping experiments use sensitive receivers to map diffuse emission. In contrast, a survey to resolve the individual sources of emission requires significantly higher sensitivity and resolution, both of which drive costs or reduce scope. Intensity mapping only requires resolution to reach cosmologically interesting scales, or scales with large enough fluctuations to secure a detection. Additionally, an intensity mapping survey is sensitive to the integral of the luminosity function, taking advantage of all the emitted radiation that is available, not just the brightest sources. However, the lack of source discrimination also makes intensity mapping survey volumes significantly more difficult to interpret than spectroscopic galaxy surveys. An intensity survey generally has not just the line emission but also all other sources of continuum and transition radiation from other redshifts.

This paper describes a method for estimating the power spectrum of intensity mapping volumes, subject to bright foreground emission and instrumental response. We inherit from the framework \citep{1997PhRvD..55.5895T, 1998ApJ...499..555T} of optimal quadratic estimators and extend methods that can be applied to single-dish intensity surveys \citep{2011PhRvD..83j3006L, 2013PhRvD..87d3005D, 2014PhRvD..89b3002D, 2013MNRAS.434L..46S}. Rather than using a fully optimal estimator, we construct a more generic estimator and new methods for handling the impact of the instrumental beam and foreground cleaning. Treatment of foregrounds is our primary focus, which in a nutshell translates into specifying the most effective foreground covariance matrix to down-weight contamination. We argue that the data themselves are the best source of information about foreground covariance, especially in light of the instrument's response to bright foregrounds.

Intensity mapping was originally developed for $21$\,cm radiation \citep{1979MNRAS.188..791H, 1990MNRAS.247..510S, 1997ApJ...475..429M}, but has been studied for several other lines (in increasing frequency): deuterium \citep{2006PhRvL..97i1301S}, $^3{\rm He}+$ \citep{2009PhRvD..80f3010M, 2009arXiv0905.1698B}, CO \citep{2008A&A...489..489R, 2011ApJ...730L..30C, 2011ApJ...741...70L, 2014MNRAS.443.3506B, 2015arXiv150308833L}, \CII\ \citep{2012ApJ...745...49G, 2014arXiv1410.4808S, 2014ApJ...793..116U, 2015arXiv150406530Y}, ${\rm Ly}\alpha$ \citep{2014ApJ...785...72G, 2014ApJ...786..111P}, C and O fine-structure \citep{2012MNRAS.419..873K}, and X-ray lines \citep{2012A&A...547A..21H}. 
 
Many experiments to search for redshifted line emission are planned or under way. These experiments deploy a wide range of technologies to observe across the frequency range of the lines of interest, from the present to the dark ages. We will focus specifically on non-interferometric, ``single-dish" methods that have a common set of simpler instrumental considerations. ``Single dish" will refer to any single-aperture optical path, including refractive designs. Throughout, the beam or point-spread functions are constant in time, axisymmetric, frequency dependent and may also have some off-diagonal Mueller mixing. While there are many parallels in interferometers, they are beyond the scope of this work (see, e.g. \citet{2014PhRvD..89b3002D, 2014ApJ...781...57S, 2014arXiv1401.2095S, 2015arXiv150206016A}). 

Aside from the $21$\,cm transition, most proposed or active intensity mapping experiments use a single-dish architecture. CO has been sought by \citet{2013ApJ...768...15P} and proposed by COMAP \citep{2015arXiv150308833L}. TIME \citep{2014SPIE.9153E..1WC} and SPHEREX \cite{2014arXiv1412.4872D} are proposed for \CII\ and ${\rm Ly}\alpha$ respectively. Within $21$\,cm efforts, single-dish instruments have been used (GBT, \citet{2010Natur.466..463C, 2013ApJ...763L..20M, 2013MNRAS.434L..46S}) or proposed (BINGO, \citet{2012arXiv1209.1041B}) for studies at $z\sim 1$. The methods described here were developed for GBT studies. Even at low redshifts, $21$\,cm interferometers such as BAOBAB \citep{2013AJ....145...65P} and CHIME \citep{2014SPIE.9145E..22B} are needed to compete with dark energy constraints from optical galaxy surveys. Interferometers are the only realistic methodology for $21$\,cm studies of reionization. 

Diffuse radiation from cosmologically redshifted atomic transitions has only been conclusively detected in cross-correlation with spectroscopic galaxy surveys. The cross-power with a density field is not biased by foregrounds, which instead boost errors (assuming that foregrounds are unrelated to signal). \citet{2015arXiv150404088C} recently detected ${\rm Ly}\alpha$ emission intensity in cross-correlation with BOSS quasars from $z=2-3.5$.  \citet{2013ApJ...763L..20M} detected $21$\,cm radiation at $z\sim 1$ in cross-correlation between dedicated GBT observations and the WiggleZ survey \citep{2010MNRAS.401.1429D} and inferred the $21$\,cm contribution to the auto-power \citep{2013MNRAS.434L..46S}. Bounds on the auto-power at modest redshift date to \citet{1986MNRAS.218..577B}. In the absence of a coeval spectroscopic galaxy survey, such as at reionization, cross-correlation with intensity maps of other atomic lines (e.g. \citet{2010JCAP...11..016V}) could secure a detection of cosmological structure, up to challenges of correlated foregrounds. As we will argue below, the principal challenge of intensity mapping experiments is that line radiation only makes up $\sim 10^{-2}-10^{-3}$ of the intensity of fluctuations in continuum radiation at most frequencies of interest.

The optimal estimator for the power spectrum requires the covariance matrix of the maps, and specification of this covariance is a central challenge of analysis of intensity mapping data. While the non-Gaussianity of foregrounds could be distinguishable from the near-Gaussianity of the signals, we will not consider separation using higher-point statistics. The full foreground covariance of a 3D survey is an $N_{\rm pix} \times N_{\rm pix}$ matrix for $N_{\rm pix}$ total map pixels and hence already requires an enormous amount of information about foregrounds and the instrument response to foregrounds. We have little prior knowledge of either.

Common approaches to specifying foreground covariance amount to different forms of dimensionality reduction. \citet{2011PhRvD..83j3006L} show that most of the covariance that distinguishes foregrounds is in the $\nu,\nu'$ directions rather than combinations involving angular separations. Fitting polynomials along the lines of sight corresponds to an ansatz of $\nu,\nu'$-only covariance contributed by those polynomial modes (e.g. \citet{2006ApJ...650..529W}). These ansatzes can be better tuned to astrophysical foregrounds by using models of the emission. Astrophysical synchrotron intensity is thought to be described by a limited number of spectral modes (e.g. \citet{2012MNRAS.419.3491L}). 

If foregrounds are bright and the instrument response is not sufficiently well understood, then misspecification of the foreground covariance can result in significant contamination in the maps that is not down-weighted.
For foregrounds $10^3$ times the signal, an unattributed $1\%$ error in calibration at one frequency could result in contamination that is 10 times larger than the signal. Worse still, if the spectral calibration varies in time, then each line of sight effectively sees a different bright foreground. These may even form a complete basis of bright spectral modes, making the signal indistinguishable. Faraday rotation through polarization leakage into intensity in radio surveys is another example \citep{2013ApJ...769..154M} where additional spectral modes can be produced by the instrumental response.

The position we take in this paper is that (1) some best-effort calibration has been applied, but that there is residual structure in the map related to the instrument, and (2) the intrinsic foreground cannot be modeled well in advance. In the regime of high foregrounds, measurements of foregrounds from other instruments or wavelengths may not have the fidelity to be useful for foreground subtraction. Intensity mapping surveys will often be the deepest surveys available in a region. This is ultimately due to the dimness of the atomic radiation, but another factor is that many 2D surveys stop integrating beyond the confusion limit. A typical intensity mapping experiment can benefit from thermal noise levels well below spatial confusion. These factors argue that the intensity survey volumes will be the best sources of information about foreground covariance rather than prior models of the instrument or intrinsic foreground emission.

Even if the relevant foreground covariance is separable as $\nu,\nu'$ blocks, we are unlikely to estimate that covariance matrix at full rank from independent sight lines in the data. The final dimensionality reduction we assume is that the $\nu,\nu'$ covariance estimated from the data will have a few dominant eigenvectors that are measured with high signal-to-noise ratio, while the remaining data are dominated by the cosmological signal and thermal noise. The foreground eigenvectors are spectral degrees of freedom that can be projected out of each line of sight. Determination of contaminated modes in the data themselves has been exploited in GBT data \citep{2010Natur.466..463C, 2013ApJ...763L..20M, 2013MNRAS.434L..46S}, GMRT \citep{2013MNRAS.433..639P}, and most recently in PAPER (e.g. \citet{2015arXiv150206016A}). Blind methods have been considered for SKA \citep{2014MNRAS.441.3271W, 2015MNRAS.447..400A} and BINGO \citep{2015arXiv150704561B}. \citet{2014ApJ...793..102S} develop a similar method for monopole signals. The success of this method relies on (1) whether instrument response to bright foregrounds can be explained by fewer spectral modes than the cosmological signal and (2) whether these modes can be determined from the data themselves, independently from the signal. These requirements relate to the rank of foregrounds rather than spectral smoothness.

The first requirement is intuitive and says that the signal must have degrees of spectral variation that are orthogonal to the foreground modes that are removed. The second requirement arises from the fact that the foreground modes are determined from the data themselves, which have foreground, signal, and noise. Spurious correlations between foreground and signal result in residual covariance in the maps that is anticorrelated with the cosmological signal. This general effect is familiar from the ILC bias (see, e.g. \citet{2009MNRAS.397.1355E}). Especially on scales that are large compared to the survey volume, there are too few realizations of signal fluctuations for the spurious correlation with the foregrounds to ``average down." In this case, it is difficult to disentangle signal from foreground. 
 
Section~\ref{sec:fgandgbt} briefly reviews continuum foreground levels for several transitions of interest and describes data from the GBT-wide survey \citep{2013MNRAS.434L..46S} that we use as an example throughout. Section~\ref{sec:quadest} develops the general (potentially suboptimal) quadratic estimator, the skeleton prescribed by the optimal estimator, and ways of calibrating the analysis using Monte Carlo simulations. Section~\ref{sec:fgcleaning} builds up the formalism of mode removal, line-of-sight cleaning, and rules of thumb for the impact of cleaning known modes. Rather than known modes, spectral contamination can be determined from the data themselves (Section~\ref{sec:empclean}), leading to modifications for rules of thumb of signal loss and estimation of transfer functions. Finally, Sections ~\ref{sec:finalprod} and \ref{sec:disc} describe the procedure for assembling final power spectral estimates using subseason cross-powers, weighted 2D to 1D averaging, and development of errors.
 
\section{Continuum Foregrounds and Data Used}
\label{sec:fgandgbt}

\subsection{Review of Continuum Foreground Levels}
\label{ssec:fgest}

This section briefly reviews the literature and estimates magnitudes of continuum foregrounds. The goal is not to make a precise determination, but instead to argue that intensity mapping has a common set of challenges. In detail, foreground challenges will vary across the bands and survey strategies. For example, an experiment that needs wide sky areas may not be able to avoid bright extragalactic sources, galactic emission, or zodiacal light, while small areas can be better tuned. To get a rough understanding of irreducible continuum levels, we will consider fluctuations in extragalactic radiation on angular scales where the signal has order-unity fluctuations. Here the mean line emission intensity serves as a proxy. Smaller angular scales may have higher signal variance, but a study of signal vs. continuum emission on different angular scales is deferred to future work for particular lines and redshifts.

Considering $21$\,cm first, \citet{2013ApJ...763L..20M} show intensity maps at $800$\,MHz ($z\approx 0.8$) with foreground fluctuations of order a kelvin, while the signal fluctuations are $\sim 0.2$\,mK. While the $21$\,cm reionization signal is brighter, the foregrounds are commensurately brighter because of the synchrotron spectral index, yielding a similar challenge. Considering CO(1-0) at $115$\,GHz, \citet{2014MNRAS.443.3506B} and \citet{2015arXiv150308833L} suggest mean temperatures of $\sim 1\,\mu{\rm K}$ at $z\sim 3$. The scales of interest at tens of arcminutes are reasonably analogous to the Cosmic Background Imager (CBI) \citep{2003ApJ...591..556P}, which finds fluctuations in the raw maps at the level of several hundred $\mu{\rm K}$ at $30$\,GHz. (Note that some sources could be cleaned or masked based on catalogs, but cosmic microwave background [CMB] would remain.) By $z\sim 8$, extragalactic and galactic synchrotron become more problematic, while the mean brightness is expected to be a similar order of magnitude \citep{2011ApJ...741...70L}. 

Moving to \CII\ ($157.7\,\mu{\rm m}$) at reionization, \citet{2014arXiv1410.4808S} estimate a mean intensity of $4 \times 10^2$\,${\rm Jy}\,{\rm sr}^{-1}$ for $z=5.3-8.5$ ($300$--$200$\,GHz), while the extragalactic fluctuations on scales of several arcmin in the Atacama Cosmology Telescope (ACT) \citep{2013ApJ...762...10D} and South Pole Telescope (SPT) \citep{2011ApJ...743...90S} data in this band are $\sim 100\,\mu{\rm K}$ at $220$\,GHz, or $\sim 5 \times 10^4$\,${\rm Jy}\,{\rm sr}^{-1}$ (CMB dominated). \CII\ emission is thought to reach a maximum at $z\approx 1$, at $\sim 5\times 10^3$\,${\rm Jy}\,{\rm sr}^{-1}$ \citep{2014ApJ...793..116U}, while {\it Herschel} ATLAS \citep{2010PASP..122..499E} shows cosmic infrared background fluctuations ranging over $\sim 2 \times 10^6$\,${\rm Jy}\,{\rm sr}^{-1}$ at $350\,\mu{\rm m}$. 

For ${\rm Ly}\alpha$ ($10.1$\,eV) mapping, \citet{2015arXiv150404088C} measure the mean surface brightness in cross-correlation between quasars and spectra, finding $\nu I_\nu = (0.74 \pm 0.17) \,{\rm nW}{\rm m}^{-2}{\rm sr}^{-1}$ (at $4500 \AA$) across $z=2-3.5$, $\sim 21-35$ higher than previously expected (see, e.g. \citet{2014ApJ...786..111P}). These redshifts span the expected peak of emission from high star formation rates. Reported mean backgrounds in the BOSS spectra include all sources of radiation, including terrestrial, and are $\nu I_\nu \sim 50 \,{\rm nW}{\rm m}^{-2}{\rm sr}^{-1}$. For lower redshifts, GALEX \citep{2010ApJ...724.1389M} data suggest astrophysical backgrounds of $\sim 10\,{\rm nW}{\rm m}^{-2}{\rm sr}^{-1}$ for $5.1-8.4$\,eV, similar to the general cosmic optical background \citep{2001ARA&A..39..249H}. In the reionization era where ${\rm Ly}\alpha$ has shifted to $\sim \mu{\rm m}$, \citet{2008ApJ...683..585L} find that the extragalactic mean contribution at $3.6\,\mu{\rm m}$ is $9\,{\rm nW}{\rm m}^{-2}{\rm sr}^{-1}$. 

In summary, typical continuum contamination to intensity surveys is $10^2-10^3$ times the line contribution. The underlying challenge is the small fraction of total luminosity emitted through line radiation. Instrumental response needs to be controlled at a subpercent level, commensurate with the brightness of the foregrounds. 

We will consider emission of a single line rather than a more general SED \citep{2014arXiv1403.3727D} and neglect interlopers at other redshifts (e.g. \citet{2015arXiv150305202B}), which have received more attention than instrumental response to bright continua.

\subsection{Data Used to Demonstrate the Method}

We will use data from the GBT-wide survey to give context to the estimator described here. Previous publications review the observations and describe the power spectrum of GBT data \citep{2013MNRAS.434L..46S} and the cross-correlation \citep{2013ApJ...763L..20M} with the WiggleZ survey \citep{2011MNRAS.415.2876B}. No results from new data are reported. \citet{2013ApJ...763L..20M} describe the observations in more detail. The GBT-wide intensity survey used the prime-focus receiver to map a $\sim 7^\circ \times 4.3^\circ$ region from $700$ to $900$\,MHz with FWHM $\sim 0.3^\circ$ and 256 spectral bands. 

Our starting point will be the map that has been estimated from time-ordered data. A framework for estimating maps is well established from CMB analysis and depends on particulars such as noise correlations and frequency masking. The details of the calibration, radio frequency interference (RFI) mitigation, and mapmaking used to produce the maps used here can be found in \citet{2013PhDT.......570M}. Section~\ref{ssec:GBTcal} describes the calibration of GBT data in regard to the spectral structure of contaminated modes.

We use Gaussian signal realizations of the Empirical-NL model of \citet{2011MNRAS.415.2876B}, which uses HALOFIT \citep{2003MNRAS.341.1311S} for nonlinear power, Kaiser redshift distortions, and streaming of the Lorentz form with $\sigma_v = 300\,h\,{\rm km/s}$. To agree with Empirical-NL, we use $\Omega_m = 0.27$, $\Omega_\Lambda = 0.73$, $\Omega_b/\Omega_m = 0.166$, $h = 0.72$, and $n_s = 0.96$ from \citet{2009ApJS..180..330K}. These parameters are also used to translate the observed regions into comoving Cartesian coordinates. We use approximations to the growth factor from \citet{2010arXiv1012.2671K}. The brightness temperature of the $21$\,cm line is taken to be
\begin{equation}
T_b(z) = T_o \frac{\Omega_{HI}(z)}{10^{-3}} \left [ \frac{\Omega_m + \Omega_\Lambda (1+z)^{-3}}{0.29} \right ]^{-1/2} \left [ \frac{1+z}{2.5} \right ]^{1/2},
\end{equation}
with $T_o = 0.39\,{\rm mK}$.

\section{The Quadratic Estimator}
\label{sec:quadest}

\subsection{Map Notations}
\label{ssec:mapdef}

The intensity survey produces maps at several frequencies. We can represent these maps as a matrix $\mat{X}$ with dimensions  $N_\nu \times N_\theta$, where $N_\nu$ is the number of frequency slices and $N_\theta$ is the number of angular pixels observed. This is a stack of 2D maps, where the map at each frequency is unraveled into an $N_\theta$-long vector. An alternative is to unravel the entire 3D volume into a single vector $\vect{x}$ of length $N_\nu \cdot N_\theta$. These representations are related through the operation $vec(\mat{X}) = \vect{x}$ which unravels the stacked map matrix into a vector. The matrix form $\mat{X}$ has the useful property that operations can act explicitly on either the frequency or angular side as $\mat{A} \mat{X} \mat{B}$, for $\mat{A}$ and $\mat{B}$ here, respectively. While this looks like a quadratic conjugation of $\mat{X}$, it is linear in $\vect{x}$ through the relation
\begin{equation}
vec(\mat{A} \mat{X} \mat{B}) = (\mat{B}^T \otimes \mat{A}) \vect{x},
\end{equation}
where $\otimes$ is the Kronecker product. Another perspective is that if a matrix $\mat{W}$ multiplying $\vect{x}$ can be written separably as $\mat{W}_\nu \otimes \mat{W}_\theta$, then those weights can act independently on the spectral and spatial axes of $\mat{X}$. We will use the $\mat{X}$ representation of the data when it is convenient to call out this separable form, and $\vect{x}$ when we want a simple ``cascaded" set of operations to apply to the full map. Note that 2D intensity surveys directly analogous to CMB cross-correlation can also be pursued (see, e.g. \citet{2013ApJ...768...15P}); however, all cases studied here will be 3D.

\subsection{Quadratic Estimators}
\label{ssec:quadintro}

In this section we review the general quadratic estimator, following early work \citep{1997PhRvD..55.5895T} and recent applications to intensity mapping \citep{2011PhRvD..83j3006L, 2012MNRAS.419.3491L, 2013PhRvD..87d3005D, 2014PhRvD..89b3002D, 2013MNRAS.434L..46S}. We break with tradition somewhat by deriving expressions for a general quadratic estimator rather than specializing to the optimal case.

Take a covariance model with Gaussian thermal noise $\mat{N}$ (initially ignoring foregrounds) and signal that is decomposed as a set of amplitudes $p_\alpha$ times the covariance $\mat{C}_{,\alpha}$ of modes in the map. Throughout, commas will denote derivatives. Then
\begin{equation}
\mat{C} = \langle \vect{x} \vect{x}^T \rangle = \mat{N} + \sum_\alpha p_\alpha \mat{C}_{,\alpha}.
\label{eqn:covmodel}
\end{equation}
We will develop a particular $\mat{C}_{,\alpha}$ for $(k_\perp, k_\parallel)$ modes of the power spectra in Section~\ref{ssec:optimalchoice}.

Our goal is to estimate the amplitudes $p_\alpha$ of the covariance from a given map $\vect{x}$, and infer their errors and correlations. A general class of covariance estimators forms a quadratic combination of the data, subtracts a bias $b_\alpha$ and then takes a linear combination of the $\vect{\hat q} |_\alpha = \hat q_\alpha$, as 
\begin{eqnarray}
\hat q_\alpha &=& \vect{x}^T \mat{Q}_\alpha \vect{x} - b_\alpha \nonumber \\
\vect{\hat p} &=& \mat{R} \vect{\hat q}.
\label{eq:basicquadratic}
\end{eqnarray}
The expectation value of the quadratic combination is
\begin{equation}
\langle \vect{x}^T \mat{Q}_\alpha \vect{x} \rangle = Tr(\mat{C} \mat{Q}_\alpha)
\end{equation}
and is sensitive to the variance of the noise $\mat{N}$ as well as the signal $\sum_\alpha p_\alpha \mat{C}_{,\alpha}$. By choosing to subtract a noise bias $b_\alpha = Tr(\mat{N}\mat{Q}_\alpha)$ based on a model for $\mat{N}$, $\hat q_\alpha$ measures just the signal covariance. Finally, the matrix $\mat{R}$ takes a linear combination of the band powers $\hat q_\alpha$ to form a final estimate $\hat p_\alpha$. The vector $\vect{\hat q}$ contains pseudo-powers in the language of \citet{2002ApJ...567....2H}. While it has mainly been described in a role of decorrelating \citep{2000MNRAS.312..285H} band powers, the final linear combination of band powers by $\mat{R}$ performs several roles as follows: (1) a normalization to ensure that $\mat{Q}_\alpha$ recovers an unbiased estimate of the signal (neglecting beam and foreground considerations), (2) a correction for signal attenuation from beam convolution and foreground down-weighting, and (3) an operation that decorrelates band powers.

In the standard treatment of quadratic estimators (see, e.g. \citet{1997PhRvD..55.5895T}), one seeks to minimize the variance of the estimator $\vect{\hat p}$ subject to the Lagrange constraint that it is an unbiased estimate of true $\vect{p}$. Most literature develops expressions henceforth assuming an optimal estimator. In the case of the optimal estimator, the Fisher matrix is ubiquitous because the estimator can saturate the Cramer-Rao bound for Gaussian fields. In contrast, we will assume that $\mat{Q}_\alpha$ is given and will in general be suboptimal. Section~\ref{ssec:optimalchoice} derives the form of the optimal estimator to develop some intuition for good suboptimal estimators. The fully optimal estimator requires a complete model of the covariance for its optimal weights. In any near-term intensity mapping applications, both the signal covariance and the foreground covariance should be assumed to be unknown. Thermal noise of the instrument can be measured well and is the only prior input to the estimator. We will develop expressions for generic $\mat{Q}_\alpha$ that will be tuned to be more robust to these unknowns.

With the choice $b_\alpha = Tr(\mat{N}\mat{Q}_\alpha)$, the expectation value of our estimator for the covariance model of Equation~\ref{eqn:covmodel} is 
\begin{eqnarray}
\langle \hat q_\alpha \rangle &=& \sum_\beta p_\beta Tr(\mat{C}_{,\beta} \mat{Q}_\alpha)  \nonumber \\
\langle \vect{\hat p} \rangle &=& \mat{R} \langle \vect{\hat q} \rangle = \mat{R} \mat{M} \vect{p} = \mat{W} \vect{p}, \label{eq:expbasicquad}
\end{eqnarray}
where we have identified $\mat{M} |_{\alpha \beta} \equiv Tr(\mat{C}_{,\beta} \mat{Q}_\alpha)$ as a mixing matrix and $\mat{W} \equiv \mat{R} \mat{M}$ as the bandpower window functions. The origin of the mixing matrix is familiar from quadratic methods such as \citet{2002ApJ...567....2H} and is due to the correlation of Fourier modes sampled over a finite area (or alternately not having an orthonormal basis in a restricted survey area). 

The estimator $\vect{\hat p}$ is then a window $\mat{W}$ that weights several modes of the true, underlying signal covariance amplitudes $\vect{p}$. The matrix $\mat{M}$ is fully dictated by the estimator, but $\mat{R}$ must be chosen. One option is to pick $\mat{R}$ so that $\mat{W}_{\alpha \alpha} = 1$, ensuring that $\hat p_\alpha$ is a unit multiple of $p_\alpha$. This does not mean that $\langle \hat p_\alpha \rangle =  p_\alpha$ because $\hat p_\alpha$ will generally be a combination of several band powers. Another choice is to pick $R_\alpha$ to give a weighted average of band powers as a window function
\begin{equation}
\sum_\beta \mat{W} |_{\alpha \beta} = 1. 
\label{eqn:windowconst}
\end{equation}
This constraint does not fully specify $\mat{R}$, but we can impose an additional constraint for simplicity that $\mat{R}$ is diagonal and $\mat{R} |_{\alpha \alpha} = R_\alpha$. This matrix normalizes each band power but does not apply any sense of decorrelation, which is developed as a final step in Section~\ref{eq:errmodel}. In this case, the full estimator is 
\begin{equation}
\hat p_\alpha = R_\alpha (\vect{x}^T \mat{Q}_\alpha \vect{x} - b_\alpha),
\end{equation}
and the constraint in Equation~\ref{eqn:windowconst} fixes
\begin{equation}
R_\alpha = \left [ \sum_\beta Tr(\mat{C}_{,\beta} \mat{Q}_\alpha) \right ]^{-1}.
\label{eqn:rotenorm}
\end{equation}

We will refer to $\mat{R}$ as the ``rote" normalization because it handles any arbitrary multipliers of the quadratic data combination. For example, if we scale all the estimators $\mat{Q}_\alpha$ by $\gamma$, then $R_\alpha \propto 1/\gamma$, correcting for the normalization. Pencil-beam surveys and slabs (one beam wide in a slice many beams long) will have significant mode coupling and may warrant a different type of analysis. Here we consider application to map regions with essentially uniform sky coverage, such as ACT \citep{2013ApJ...762...10D} and SPT \citep{2011ApJ...743...90S}.

\subsection{The Optimal Estimator and Its Normalization}
\label{ssec:optimalchoice}

So far, $\mat{Q}_\alpha$ has organized a generic quadratic combination of the data. In this section, we review the formally optimal estimator and the procedure it implies for analyzing the data. The optimal estimator provides a good starting point for constructing estimators, but subsequent statements will leave general $\mat{Q}_\alpha$ rather than assume optimality. The covariance of the $\vect{\hat q}$ values in Equation~\ref{eq:basicquadratic} is
\begin{equation}
{\rm Cov}(\hat q_\alpha, \hat q_\beta) = Tr(\mat{C} \mat{Q}_\alpha \mat{C} \mat{Q}_\beta). 
\end{equation}
Minimizing the covariance results in 
\begin{equation}
\mat{Q}_\alpha = \frac{\mat{C}^{-1} \mat{C}_{,\alpha} \mat{C}^{-1}}{Tr(\mat{C}^{-1} \mat{C}_{,\alpha} \mat{C}^{-1} \mat{C}_{,\alpha})},
\end{equation}
where the normalization is fixed by the Lagrange multiplier that forces the estimator to be unbiased. 

Based on the previous section, the mixing matrix of the optimal estimator is
\begin{equation}
\mat{M} |_{\alpha \beta} = Tr(\mat{C}^{-1} \mat{C}_{,\alpha} \mat{C}^{-1} \mat{C}_{,\beta}),
\end{equation}
which is just the Fisher matrix $\mat{F}$ of the estimator (e.g. \citet{1997ApJ...480...22T}). 

This is the starting point for more robust estimators. Note that $\mat{Q}_\alpha$ is only optimal when $\mat{C}$ is known exactly. When it is not known perfectly, the estimator is generally suboptimal but can be constructed to be unbiased. 

\subsection{Optimal Estimation Procedure}
\label{ssec:optestproc}

To get intuition for the term $\mat{C}_{,\alpha}$, note that in CMB analysis it is the outer product of spherical harmonics 
\begin{equation}
\mat{C}_{,\alpha} = \vect{y}_\alpha \vect{y}_\alpha^T~~~{\rm where}~~~\vect{y}_\alpha |_i = Y_{\ell m(\alpha)} (\vect{\hat r}_i).
\label{eq:cmbsigcov}
\end{equation}
Hence, for the CMB, $\hat q_\alpha$ has the interpretation of the amplitude of the spherical harmonic transform of the weighted maps, reinforcing $\vect{\hat q}$ as a pseudo-power as in \citet{2002ApJ...567....2H}.

The operation $\mat{C}_{,\alpha}$ is more complex in intensity mapping surveys because of the relation between the geometry of the observations and the desired 3D power spectrum. $\vect{x}$ represents the calibrated intensity as a function of angular pointing and frequency and can be interpolated onto a cube in units of $h^{-1} {\rm Mpc}$ using a fiducial cosmology. Here angular slices translate approximately into spatial slices, and frequencies translate into distances to redshifts. The relation between redshift and distance is nonlinear, and the low-frequency end of the survey represents a larger spatial area than the high-frequency side. The survey is therefore approximately a truncated pyramid in comoving Cartesian coordinates. Any areas outside the pyramid can be given zero weights in the larger coordinate volume. If the angular region is large enough, there will be non-flat sky curvature in the spatial slices. Through these factors, there is not a 1-to-1 translation between observation and comoving Cartesian coordinates, and interpolation inevitably leads to loss in fidelity. Consider linear interpolation schemes that can be represented as $\mat{P} \vect{x}$, where $\mat{P}$ moves a map $\vect{x}$ to Cartesian coordinates. It is not generally invertible, but so long as the discretization is fine enough, the observed $\theta-\nu$ space maps can be translated to Cartesian and back again with little loss.

\begin{figure}[htb]
\epsscale{0.5}
\includegraphics[scale=0.6]{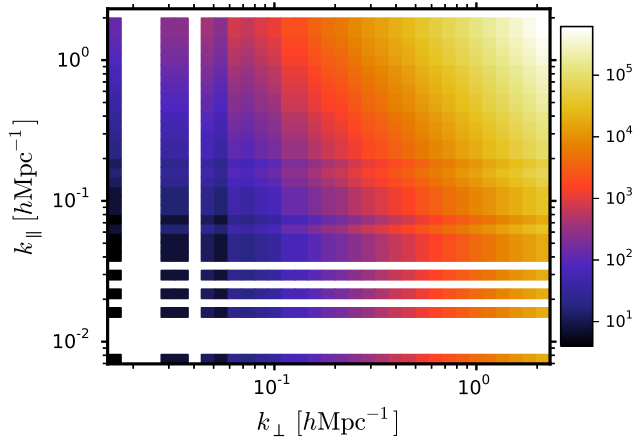}
\caption{Number of 3D Fourier modes contributing to a given band power in the GBT-wide survey. The survey region is $7^\circ \times 4.3^\circ$ from $700$ to $900$\,MHz in 256 spectral bins. Toward increasing $k_\perp$ and $k_\parallel$, the number of modes grows quadratically and linearly, respectively. The survey volume encompasses several fundamental Fourier modes that appear as isolated bands. In subsequent plots, we ignore these lone harmonics. Even though the number of available modes rapidly increases toward high $k$, information from $k_\perp > 0.4~h{\rm Mpc}^{-1}$ is almost entirely suppressed by the GBT beam.
\label{fig:counts_2dk}}
\end{figure}

Once in a grid of constant Cartesian dimensions, $\mat{C}_{,\alpha}$ can be understood as an operation that takes the 3D Fourier transform of both maps \citep{2013PhRvD..87d3005D} and bins the k-vectors into annuli in a range of $k_{\perp, \alpha}$ to $k_{\perp, \alpha+1}$ and $k_{\parallel, \alpha}$ to $k_{\parallel, \alpha+1}$ that define the range of the 2D band power $p_\alpha$. Let $\mat{K}$ be the linear Fourier operation so that $\vect{\tilde x} = \vect{K} \mat{P} \mat{C}^{-1} \vect{x}$ is the Fourier transform of the data in Cartesian coordinates. Then the binning operation is mathematically 
\begin{equation}
\hat q_\alpha = \sum_{\vect{k}} I_{\vect{k} \in A_\alpha} \vect{\tilde x} (\vect{k}) \vect{\tilde x} (\vect{k})^* \biggl / \sum_{\vect{k}} I_{\vect{k} \in A_\alpha},
\end{equation}
where $I_{\vect{k} \in A_\alpha}$ is the indicator that is 1 in the $k$-bin annulus $A_\alpha$ and 0 elsewhere. 
Figure~\ref{fig:counts_2dk} shows the number of 3D Fourier modes contributing to 2D band powers with logarithmic spacing. This operation performs no weighting, assuming that $\mat{C}^{-1} \vect{x}$ has noise isotropic in the $k_\perp$ annulus. The estimator has the form of an inner product $\vect{\tilde x}^T \mat{B}_\alpha \vect{\tilde x}$ where $\mat{B}_\alpha$ performs the binning of 3D Fourier cells. Combined, $\mat{C}_{,\alpha} = \mat{P}^T \mat{K}^T \mat{B}_\alpha \mat{K} \mat{P}$ and can be understood as taking the Fourier transform of both maps in Cartesian coordinates and then binning onto the band power $\alpha$. This can be easily parallelized across band powers by using the same Cartesian space conversion and Fourier transform. Note that $\mat{C}^{-1}$ is applied to the maps on both sides in observing coordinates of $\theta$ and $\nu$. It will remain natural to discuss covariance in those dimensions rather than Cartesian coordinates because contamination naturally lives along $\nu$.

The matrix calculations throughout this paper are skeletons that put the proper form to procedures implemented in software. Putting together what we have so far, $\vect{x}^T \mat{Q}_\alpha \vect{x}$ can be written as a numerically convenient procedure:
\begin{enumerate}
\item Weight both maps by their inverse covariance in observing coordinates (and remove the mean map if needed).
\item Translate observing to Cartesian coordinates.
\item Calculate the fast Fourier transform to both sides of the quadratic product.
\item Bin onto a band power.
\end{enumerate}

When considering one Fourier mode, the binning operation is separable \citep{2014PhRvD..90b3019L} as $\mat{B}_\alpha = \mat{D}^T \mat{D}$. In this case, it is useful to derive the map in observing coordinates that has information about a Fourier mode. Doing this requires transforming through $\mat{P}$ and back through $\mat{P}^{-1}$, which we assume to exist even though information could generally be lost in the repixelization. This ``filter" for the modal information in a map becomes $\vect{x}_\alpha' = \mat{P}^{-1} \mat{K}^T \mat{D}_\alpha \mat{K} \mat{P} \mat{C}^{-1} \vect{x}$. The map $\vect{x}_\alpha'$ is the Fourier component $\alpha$ of the noise-weighted input map, and the complete quadratic estimator is simply $\vect{x}_\alpha'^T \vect{x}_\alpha'$. In later sections, we will estimate rules of thumb for the impact of foreground operations on signal by considering these simple forms of inner products.

\subsection{Estimating the Normalization}
\label{ssec:simpath}

There is complete freedom in choosing $\mat{R}$ in Equation~\ref{eq:expbasicquad}. For arbitrary $\mat{R}$, $\mat{W} = \mat{R} \mat{M}$ translates theoretical expectations $\vect{p}$ to the same space as the measured data $\vect{\hat p}$. If $\mat{R}$ is invertible, there is no information loss, and the choice of $\mat{R}$ relates to the presentation of data. For example, we argued that the choice of Equation~\ref{eqn:rotenorm} corrected for some multiplier $\gamma$ in the estimator, but there is no reason it needs to. The theoretical $P(k)$ could be compared to the data through the same pipeline with the arbitrary $\gamma$ retained, effectively comparing to $\gamma P(k)$. A non-invertible $\mat{R}$ would be a poor choice because information is lost, but any information that does get through would permit a comparison of theory and observation.

The choice of $\mat{R}$ is especially relevant to intensity mapping because signal is attenuated by the beam and foreground down-weighting. One outlook is that the theory $P(k)$ should have the same beam and foreground treatment applied as in the real data and then be compared to the ``raw" experimental band powers, without making any effort to correct the observed band powers. This makes it difficult to intercompare experiments with different foreground properties and beams, or even the same experiment with different foreground treatments. We prefer to bring the measured $\vect{\hat p}$ into the same footing as the theoretical band powers $\vect{p}$ from $P(k)$. This means correcting for the beam (Section~\ref{sec:beamimpact}) and foreground cleaning (Section~\ref{ssec:directloss}).

Normalizations take the form $Tr(\mat{C}_{,\beta} \mat{Q}_\alpha)$ (see Equation~\ref{eqn:rotenorm}). The previous section described a software procedure to calculate $\vect{x}^T \mat{Q} \vect{x}$, but not the trace with an arbitrary matrix. We would like to estimate the normalization using the same well-defined pipeline that calculates $\vect{x}^T \mat{Q} \vect{x}$, rather than develop a separate matrix operation. This reuse reduces the complexity of software, enforces consistency through common pipelines, and provides a convenient numerical implementation. 

$\sum_\beta Tr(\mat{C}_{,\beta} \mat{Q}_\alpha)$ is the expectation value of band powers when $p_\beta = 1$, so that we can estimate $R_\alpha$ in Monte Carlo by finding the mean $\hat q_\alpha = \vect{x}^T_{p=1} \mat{Q}_\alpha \vect{x}_{p=1}$ of white noise in the input $\vect{x}_{p=1}$ that is drawn from a covariance where $p_\beta = 1$. In other words, $\mat{R}$ ensures that unit power in is unit power out. Monte Carlo estimation is efficient because there are typically many more map pixels than estimated band powers, so a single Monte Carlo involves significant averaging. The number of samples varies with estimator, but for GBT-wide, several hundred were typically sufficient. 

Rather than lump the normalization into a single factor that is estimated with simulations, we prefer to partition the rote normalization, effects of beam convolution (Section~\ref{sec:beamimpact}), and foreground down-weighting as separate operations (Section~\ref{ssec:fgtrans}). The rote normalization will ensure that the power spectral estimator of the spatially weighted maps properly recovers inputs. The foreground transfer function accounts for attenuation of the cosmological signal by foreground down-weighting. Section~\ref{sec:fgcleaning} develops foreground treatment that is naturally partitioned into operations that avoid contaminated modes and weigh the survey based on thermal noise. Finally, the beam transfer function accounts for differences between the measured band powers and the inputs due to convolution by the instrumental response. We develop the beam transfer function first because it is simplest.  

\subsection{Impact of the Instrumental Beam}
\label{sec:beamimpact}

Diffraction limits the resolution of single-dish (aperture) and interferometer instruments (baseline). We will only consider the case of the single-dish instrument with an axially symmetric beam. Beams that are not axially symmetric have a non-isotopic impact on the data in $k$-space. Also, typical surveys cover a region at several parallactic angles, so a beam that is not axially symmetric does not correspond to the stationary convolution across the map. When the beam convolution applies uniformly across the map, it is a multiplication in Fourier space, so let a particular Fourier mode be modulated as
$\mat{C}_{,\alpha}^B = B_\alpha \mat{C}_{,\alpha}$. Throughout, the 2D band powers $\alpha$ combine the annulus in $k_x$ and $k_y$ at constant $k_\perp = \sqrt{k_x^2 + k_y^2}$, so $B_\alpha$ already implicitly assumes an axisymmetric beam. The expectation value of the estimator is 
\begin{eqnarray}
\langle \hat q_\alpha \rangle &=& \sum_\beta B_\beta p_\beta Tr(\mat{C}_{,\beta} \mat{Q}_\alpha) = \mat{M} \mat{B} \vect{p} \\
\langle \vect{\hat p} \rangle &=& \mat{R}_B \langle \vect{\hat q} \rangle = \mat{R}_B \mat{M} \mat{B} \vect{p} = \mat{W}_B \vect{p}.
\end{eqnarray}
The window function $\mat{W}_B \equiv \mat{R}_B \mat{M} \mat{B}$ includes the effect of the beam, and $\mat{R}_B$ reflects that we will want a different normalization due to the impact of the beam. The convolution acts on the underlying band powers, which are then mixed under $\mat{M}$ through the estimator, identically to, e.g. \citet{2002ApJ...567....2H}. In a context like the optimal quadratic estimator, where $\mat{Q}_\alpha = \mat{C}^{-1} \mat{C}_{,\alpha}^B \mat{C}^{-1}$, the estimator itself scales with the beam. In this case, $\mat{M}$ also scales as $B_\alpha$ and the rote normalization Equation~\ref{eqn:rotenorm} scales as $B_\alpha^{-1}$, so the effect cancels. We will neglect the impact of the beam on the estimator, assuming fixed $\mat{Q}_\alpha$ for simplicity.

\begin{figure}[htb]
\epsscale{0.5}
\includegraphics[scale=0.6]{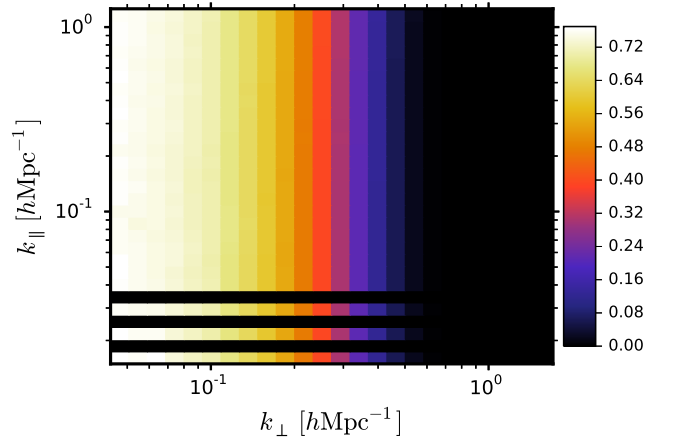}
\caption{Transfer function of the GBT beam. This is derived as the ratio of signal simulations with beam to those without. It is analogous to $B_\ell$ in CMB analysis, and here it clearly acts along the spatial ($k_\perp$) directions. Chromatic aspects of the beam are eliminated by convolving the maps to a common resolution, at the low-frequency end, with FWHM $\sim 0.3^\circ$. 
\label{fig:beam_transfer}}
\end{figure}

Choose to keep the same rote normalization of the quadratic estimator as Equation~\ref{eqn:rotenorm} but extend the transformation $\mat{R}_B$ to include a diagonal ``beam transfer function" correction for the beam $\mat{T}_B^{-1}$, which is $\mat{R}_B \equiv \mat{T}_B^{-1} \mat{R}$. The expectation value is then
\begin{equation}
\langle \vect{\hat p} \rangle =  \mat{T}_B^{-1} \mat{R}  \mat{M} \mat{B} \mat{p} = \mat{W}_B \vect{p}.
\end{equation}
The transfer beam function can be determined by enforcing $\sum_\beta \mat{W}_B |_{\alpha \beta} = 1$. Again choose a diagonal $T^B_{\alpha} = \mat{T}_B |_{\alpha \alpha}$ so that
\begin{equation}
(T^B_{\alpha})^{-1} R_\alpha \bar B_\alpha \sum_\beta Tr(\mat{C}_{,\beta} \mat{Q}_\alpha) = 1
\end{equation}
where
\begin{equation}
\bar B_\alpha = \sum_\beta Tr(\mat{C}_{,\beta} \mat{Q}_\alpha) B_\beta \biggl / \sum_\beta Tr(\mat{C}_{,\beta} \mat{Q}_\alpha).
\label{eq:meanbeam}
\end{equation}
This is the ratio of simulations with beam convolution to those without, which can be estimated using the same pipeline as the data. Plugging in the value from the rote normalization, Equation~\ref{eqn:rotenorm} yields $T^B_{\alpha} = \bar B_\alpha$. With the choices here, the transfer function is the mixing-weighted beam. Figure~\ref{fig:beam_transfer} shows the beam transfer function estimated from simulations. The full estimator has become
\begin{equation}
\hat p_\alpha = \bar B_\alpha^{-1} R_\alpha (\vect{x}^T \mat{Q}_\alpha \vect{x} - b_\alpha).
\end{equation}
This follows \citet{1997PhRvD..55.5895T} rather than \citet{2002ApJ...567....2H}, where the data are multiplied by $\mat{M}^{-1}$ and then the beam is treated. Our model throughout is to normalize the band powers (including the effect of beam and foreground) before decorrelating. Because of the effects of spatial and spectral masking and foreground deweighting, the mixing matrix of an intensity survey is not as easily calculable as the CMB. Here decorrelation is a final step related to display of the data and is based on simulations (Section~\ref{eq:errmodel}) rather than a closed-form calculation.

The beam and its sidelobes generally broaden toward lower frequencies. As the sidelobes expand over the spatially varying continuum foreground structure, they can produce spectral structure. This is analogous to the wedge phenomenon in interferometers (see, e.g. \citet{2014PhRvD..90b3019L}). In a single-dish setting with approximately uniform map coverage, we can convolve all maps to a common resolution using the beam model. Figure~\ref{fig:beam_transfer} has no structure in $k_\parallel$ because of this operation. 

\section{Cleaning Known Foreground Modes}
\label{sec:fgcleaning}

The estimator developed so far has referred to general signal and noise covariances. Thermal noise clearly belongs in the noise covariance $\mat{N}$ in Equation~\ref{eqn:covmodel}. Foregrounds are more challenging to attribute. Like thermal noise, they additively bias the power spectral estimates. Unlike thermal noise, they are not known accurately in advance, so the power spectral bias cannot be simply subtracted based on a model. They are also unavoidable astronomical signals, while thermal noise biases can be avoided by calculating the cross-power between subseasons with uncorrelated noise (as we will do in Section~\ref{ssec:crosspwr}). The perspective we take follows \citet{2013MNRAS.434L..46S} and \citet{2014PhRvD..89b3002D} by including any additive biases from residual foregrounds in the final power spectrum. This assumes that any foregrounds that can be modeled are down-weighted or subtracted, and residuals are, by definition, not possible to model and subtract as a noise bias. Residual foregrounds are indistinguishable from signal.

If foregrounds were drawn from a Gaussian distribution and known in advance, their covariance would be a complete description. This covariance would enter $\mat{C}^{-1}$ and be down-weighted in the maps. In reality, we lack detailed knowledge of how a particular experiment will respond to bright foregrounds in a particular region of the sky. Further, both extragalactic fluctuations and the galaxy are non-Gaussian, so $\mat{C}^{-1}$ will not properly down-weight the contaminated modes. A more conservative choice is to admit some generally contaminated modes that are fully projected out of the data rather than subjected to a more nuanced weight. These ``avoided" modes are easy to include in the quadratic estimator. 

Take the data to be true sky signal $\vect{x}_{\rm sig}$ plus some set of normalized modes $\mat{Z}$ multiplied by amplitudes $\vect{a}$. The form of the quadratic estimator that we seek is then 
\begin{equation}
\hat q_\alpha = (\vect{x}_{\rm sig} + \mat{Z} \vect{a})^T \mat{Q}_\alpha (\vect{x}_{\rm sig} + \mat{Z} \vect{a}) - b_\alpha,
\end{equation}
where estimators are orthogonal to $\mat{Z}$, or $\mat{Q}_\alpha \mat{Z} =0$. This choice forces $\hat q_\alpha$ to be uninfluenced by the modes in the matrix $\mat{Z}$. The quadratic estimator can be derived in the usual way of minimizing the covariance subject to the constraint that the result is unbiased (a normalization). In addition, a Lagrange multiplier can force $\mat{Q}_\alpha \mat{Z} =0$. Appendix B4 of \citet{1998ApJ...499..555T} shows that the optimal
\begin{eqnarray}
\mat{Q}_\alpha &=& \mat{\Pi}^T \mat{C}^{-1} \mat{C}_{,\alpha} \mat{C}^{-1} \mat{\Pi}, \\
\mat{\Pi} &\equiv& \mat{1} - \mat{Z}(\mat{Z}^T \mat{C}^{-1} \mat{Z})^{-1} \mat{Z}^T \mat{C}^{-1}.
\label{eq:avoidmode}
\end{eqnarray}
Note that $\mat{\Pi}$ is idempotent $\mat{\Pi} \mat{\Pi} = \mat{\Pi}$ and so is a projection of the signal onto a space orthogonal to the modes $\mat{Z}$.  

There are several equivalent viewpoints on foreground cleaning. Above, we have presented a constrained quadratic estimator that is orthogonal to contaminated modes. A Bayesian approach would jointly estimate contaminated modes with the signal modes in the map, but then marginalize over contaminated modes to recover the signal. This marginalization is functionally equivalent to the orthogonal estimator \citep{1998ApJ...503..492S}. Both of these outlooks treat the contamination as additional ``modes" to estimate, much like the signal. An alternative is to lump the contaminated modes with the noise covariance that is handled by the $\mat{C}^{-1}$ operation. This too is connected to the above approach through the Woodbury identity
\begin{equation}
\mat{C}^{-1} \mat{\Pi} = \lim_{\sigma^2 \rightarrow \infty} (\mat{C} + \sigma^2 \mat{Z} \mat{Z}^T)^{-1} = (\mat{C} + \mat{F})^{-1}.
\end{equation}

The estimator avoids the foreground modes $\mat{Z}$, so $\mat{C}$ represents thermal noise, cosmological signal, and any residual foregrounds that are not explained by the modes $\mat{Z}$.

Again there is freedom in choosing $\mat{C}$ to yield slightly less optimal but more robust estimators. We will continue to ignore the signal covariance contribution in the inverse noise weights. Until a high-significance signal detection is reached, signal covariance will be both subdominant and imprecisely known. 
We will assume that $\mat{C}$ is purely thermal noise and is diagonal. $\mat{C}^{-1}$ translates simply into weighting pixels by the inverse of the variance from thermal noise, which is related to the integration time and effective $T_{\rm sys}(\nu)$ of the survey. This prescription allows $\mat{C}$ to be precisely determined using survey properties and checked using the difference maps between subseasons (Section~\ref{ssec:crosspwr}).

We can correct for the impact of foreground cleaning in the same scheme as the beam. Let the band power estimate be
\begin{equation}
\hat p_\alpha = (T^B_\alpha T^F_\alpha)^{-1} R_\alpha (\vect{x}^T \mat{Q}_\alpha \vect{x} - b_\alpha),
\end{equation}
where $T^B_\alpha$ is the beam transfer function correction and $R_\alpha$ is the rote normalization, as before. We can form a foreground cleaning transfer function $T^F_\alpha$ using simulations that compare the ratio of the estimated band power $\alpha$ before and after foreground cleaning is applied to the data. More formally, 
\begin{equation}
T^F_\alpha = \frac{\sum_\beta Tr(\mat{C}_{,\beta} \mat{\Pi}^T \mat{C}^{-1} \mat{C}_{,\alpha} \mat{C}^{-1} \mat{\Pi})}{\sum_\beta Tr(\mat{C}_{,\beta} \mat{C}^{-1} \mat{C}_{,\alpha} \mat{C}^{-1})}.
\label{eq:fgtrans}
\end{equation}

The transfer function must be measured in 2D $k$-space rather than 1D $k$ shells because of the anisotropic nature of the cleaning. Foreground cleaning is not a real-space convolution like the beam, so is not fully described by a band power multiplier. The factors $(T^B_\alpha T^F_\alpha)^{-1} R_\alpha$ give the proper normalization but do not attempt to de-correlate the band powers. Section~\ref{eq:errmodel} describes how covariance simulations can be used to produce a final summary with independent errors per band power.

\subsection{Line-of-sight Cleaning: Known Modes}
\label{ssec:losclean}

The modes in $\mat{Z}$ could contain both spatial and spectral information. Most of the foreground variance that distinguishes it from the signal is in the $\nu,\nu'$ directions \citep{2011PhRvD..83j3006L}, so $\mat{Z}$ can be separated into spectral modes per line of sight and yield a separable projection $\mat{\Pi} = \mat{\Pi}_{\nu} \otimes \mat{1}$. Rather than identifying a handful of contaminated modes, it is convenient to work with a complete spectral foreground basis where only a handful of modes are projected out. Starting with a limited set of foreground spectral modes $\vect{u}^{\rm f}_i$, a complete basis can be constructed using a Gram-Schmidt process. (Superscripts without a numerical value are used as labels rather than powers.)
Assemble these complete basis vectors into columns of a matrix $\mat{U}_{\rm f}$ that projects onto a ``spectral foreground" basis. The matrix $\mat{X}$ representation of the map (Section~\ref{ssec:mapdef}) has dimensions $N_\nu \times N_\theta$ and can be put in this basis as $\mat{U}_{\rm f}^T \mat{X}$. This combination is a stack of maps of the amplitudes of spectral modes. The line-of-sight projection operation becomes 
\begin{equation}
\mat{\Pi} \vect{x} = vec([\mat{1} - \mat{U}_{\rm f} \mat{S} (\mat{U}_{\rm f}^T \mat{C}^{-1}_{\nu \nu'} \mat{U}_{\rm f})^{-1} \mat{U}_{\rm f}^T \mat{C}_{\nu \nu'}^{-1} ] \mat{X}),
\label{eq:weightedlos}
\end{equation}
where we have added a diagonal selection matrix $\mat{S}$ that is $1$ for those modes that are subtracted and $0$ for spectral modes that pass through. The inverse covariance in this case is $\mat{C}^{-1}_{\nu \nu'}$, which under the assumptions here is just the $\nu,\nu'$ covariance of the thermal noise. This is a diagonal matrix derived from $T_{\rm sys}(\nu)$. For notational simplicity, we will assume that the thermal noise is constant across all frequencies so that $\mat{C}^{-1}_{\nu \nu'} = \mat{1}$ and
\begin{equation}
\mat{\Pi} \vect{x} = vec((\mat{1} - \mat{U}_{\rm f} \mat{S} \mat{U}_{\rm f}^T) \mat{X}),
\label{eq:losproj}
\end{equation}
using orthonormality of the modes $\mat{U}_{\rm f}$. This has the form of a simple projection of spectral modes. Experiments whose noise varies with frequency (or has masked frequencies) should use the $\mat{C}^{-1}_{\nu \nu'}$-weighted form. Rather than carry around the full $\mat{\Pi}$, define the line-of-sight projection that acts on the left side of the map $\mat{X}$ as 
\begin{equation}
\mat{\Pi}_\nu \equiv \mat{1} - \mat{U}_{\rm f} \mat{S} \mat{U}_{\rm f}^T.
\label{eq:simplelosclean}
\end{equation}

\subsection{Cleaning Effectiveness and Direct Loss}
\label{ssec:directloss}

In this section, we use the variance as a simple measure of the impact of foreground cleaning, rather than the more complex bandpower. The variance is directly related to the eigenvalue spectrum and modal structure of signal foregrounds and provides rules of thumb and intuition about the process of foreground cleaning. Let the $\nu,\nu'$ covariance of a pure signal map $\smap$ be $\mat{C}_s = N_\theta^{-1} \langle \smap \smap^T \rangle$. Diagonalize the signal covariance as $\mat{C}_s = \sfmodes \mat{\Lambda}_s \sfmodes^T$, where $\mat{\Lambda}_s |_{ii} = \lambda^{\rm s}_i$ and $\sfmodes$ is made of the signal spectral eigenvectors $\vect{u}^s_i$. The signal variance is
\begin{eqnarray}
\xi_s &=& N_\theta^{-1} \langle \vect{x}_s^T \vect{x}_s \rangle = N_\theta^{-1} \langle Tr (\smap \smap^T) \rangle \\
&=& Tr (\sfmodes \mat{\Lambda}_s \sfmodes^T) = \sum_k \lambda^{\rm s}_k,
\label{eqn:sigcov}
\end{eqnarray}
where the expectation value is over signal realizations, and we have identified the Frobenius trace $Tr (\smap \smap^T)$.
This is the simplest quadratic estimator, corresponding to $\mat{Q} = \mat{1}$. Recall that the operation $\vect{x}_\alpha' = \mat{P}^{-1} \mat{K}^T \mat{D}_\alpha \mat{K} \mat{P} \mat{C}^{-1} \vect{x}$ developed in Section~\ref{ssec:optestproc} can filter all of the modal content of some band power $\alpha$ onto a new map $\vect{x}_\alpha'$ so that $\vect{x}_\alpha'^T \vect{x}_\alpha'$ is the full quadratic estimator.

Similar to the signal, pure foreground spectral covariance can also be decomposed into eigenmodes $\fgfmodes$ as $\mat{C}_{\rm f} = N_\theta^{-1} \langle \fgmap \fgmap^T \rangle = \fgfmodes \mat{\Lambda}_{\rm f} \fgfmodes^T$. After applying the cleaning $\mat{\Pi}_\nu$, the trace of the foregrounds becomes
\begin{equation}
\xi^{fg}_{clean} = N_\theta^{-1} Tr(\mat{\Pi}_\nu \fgmap \fgmap^T \mat{\Pi}_\nu^T) = \sum_{i \notin {\rm cuts}} \lambda^{\rm f}_i.
\label{eq:fgdirectloss}
\end{equation}
The variance of residual foregrounds in the map is the sum of the eigenvalues of the modes that were not projected out. 

The foreground cleaning operation does not null discrete signal modes because signal and foreground have a different basis. A good measure of this signal loss is to apply foreground cleaning to a signal map ($\mat{\Pi}_\nu \smap$) and find the cross-variance with the input signal map $\smap$,
\begin{eqnarray}
\xi_{\rm clean} &=& N_\theta^{-1} \langle Tr(\mat{\Pi}_\nu \smap \smap^T) \rangle = \xi_s + \xi_{\rm direct} \\
\xi_{\rm direct} &\equiv& -N_\theta^{-1}\langle Tr(\fgfmodes \mat{S} \fgfmodes^T \smap \smap^T) \rangle,
\label{eq:signallosstrace}
\end{eqnarray}
where we have identified the signal covariance from Equation~\ref{eqn:sigcov} and defined the loss of signal as a direct result of cleaning as $\xi_{\rm direct}$, which will be negative. Hence, the signal variance in the cleaned map is the input signal variance minus some loss. We will show that this loss is the overlap of signal with the foreground modes that were projected out. Writing out the direct loss,
\begin{equation}
\xi_{\rm direct} = -\sum_{j \in {\rm cuts}, k} \lambda^{\rm s}_k \langle [(\vect{u}^s_k)^T \vect{u}^{\rm f}_j]^2 \rangle,
\label{eq:xidirect}
\end{equation}
which is the overlap of the signal modes with the foreground modes that are subtracted. We can develop a simple expression by assuming that any foreground mode is uncorrelated with any random signal mode. In this case, spurious correlations scale as $(\vect{u}^s_k)^T \vect{u}^{\rm f}_j \propto 1/\sqrt{N_{{\rm res},\nu}}$, where $N_{res,\nu}$ is the number of independent modes of the signal over the frequency range. For example, at the highest $k$ in the box, this approaches $N_\nu$, while for lower $k$ there are relatively fewer spectral modes in the signal. If $N_m$ modes are removed in the sum over $j \in {\rm cuts}$,
\begin{equation}
\xi_{\rm clean} = \xi_s + \xi_{\rm direct} \sim \left (1 - \frac{N_m}{N_{{\rm res}, \nu}} \right ) \xi_s.
\label{eq:givenfgrule}
\end{equation}
Detection of signal in the cleaned maps benefits as the number of signal degrees of freedom $N_{{\rm res}, \nu}$ exceeds the number of foreground modes removed, $N_m$.

There will generally be some spurious correlation between foreground spectral modes and signal spectral modes. Under an assumption that the foregrounds are physically unrelated to the signal, there is no correlation to the signal, on average. Hence, spurious correlation is an issue of noise and signal loss rather than bias. (This can be violated when the sources of line radiation produce continuum emission, but that case is left for future work.)

In practice, we expect that intensity mapping experiments will need to measure at least some contaminated modes from the data themselves. In this case, the signal influences the foreground modes. We next show that this produces a net anticorrelation of the cleaned foregrounds with the signal, resulting in considerable bias on the largest scales in the map. This effect is familiar from ILC bias in CMB foreground cleaning (see, e.g. \citet{2009MNRAS.397.1355E}), and in general of blind methods. 

\section{Empirical Cleaning}
\label{sec:empclean}

The instrument response to bright foregrounds will not follow a Gaussian distribution or be fully known in advance. Some examples of instrument response are passband calibration, chromatic beam response, polarization leakage, and calibration stability. Experiments will make a best effort at calibrating these effects, but any differences will modulate how the experiment observes bright foregrounds. Further, we would like to avoid assuming that the intrinsic spectrum of contaminants is known in advance. 

A limited number of contaminated spectral modes can be estimated from the data themselves by finding the empirical $\nu,\nu'$ covariance $\mat{\hat C}$ and its eigenvalue decomposition $\mat{\hat C} = N_\theta^{-1} \mat{X} \mat{X}^T = \mat{\hat U} \mat{\hat \Lambda} \mat{\hat U}^T$. Unlike the previous section, we do not take the average over signal realizations through $\langle \rangle$, assuming instead that we only have access to one map $\mat{X}$ that is the sum of foregrounds, signal, and noise. Writing $\mat{\hat C}$ in terms of these constituents, 
\begin{equation}
\mat{\hat C} = N_\theta^{-1} (\fgmap + \smap + \nmap) (\fgmap + \smap + \nmap)^T.
\end{equation}
Initially neglect the noise contribution so that
\begin{eqnarray}
\mat{\hat C}_{\rm s+f} &=& N_\theta^{-1} (\fgmap \fgmap^T + \fgmap \smap^T + \smap \fgmap^T + \smap \smap^T) \nonumber \\
&=& \mat{\hat C}_{\rm f} + \mat{C}_\Delta,
\end{eqnarray}
where $\mat{C}_\Delta \equiv N_\theta^{-1} (\fgmap \smap^T + \smap \fgmap^T + \smap \smap^T)$, and $\rm s+f$ denotes the fact that the covariance contains both signal and foregrounds. Thus, the eigenvalues that we estimate from $\mat{\hat C}$ likely are dominated by foreground $\mat{\hat C}_{\rm f}$, but also perturbed by the signal itself (and noise) through $\mat{C}_\Delta$. Write the updated eigenvalues of $\mat{\hat C}$ as
\begin{equation}
\mat{U}_{\rm s+f} = \fgfmodes + \mat{\Delta}.
\end{equation}
The data-cleaning operation is then with respect to these perturbed eigenvectors, defining
\begin{equation}
\mat{\Pi}^\nu_{\rm s+f} \equiv 1 - (\fgfmodes + \mat{\Delta}) \mat{S} (\fgfmodes + \mat{\Delta})^T.
\end{equation}
This again projects a subset of contaminated modes from each line of sight. In comparison, the ideal cleaning would only remove foreground modes as
\begin{equation}
\mat{\Pi}^\nu_{f} \equiv 1 - \fgfmodes \mat{S} \fgfmodes^T.
\end{equation}

There is not a general analytic theory describing the eigenvectors of the sums of matrices. It is possible to write a rank-$N$ update of the eigenvectors that gives some intuition that the signal ``rotates" foreground modes, but that direction does not yield analytic expressions. The next section develops expressions that are perturbative in small signal, and some analytic insights are possible. Ultimately, the properties of this cleaning need to be estimated numerically in transfer function simulations (Section~\ref{ssec:fgtrans}). 

\subsection{Example: GBT Foreground Decomposition and Instrument Response}
\label{ssec:GBTcal}

\begin{figure}[htb]
\epsscale{0.5}
\includegraphics[scale=0.6]{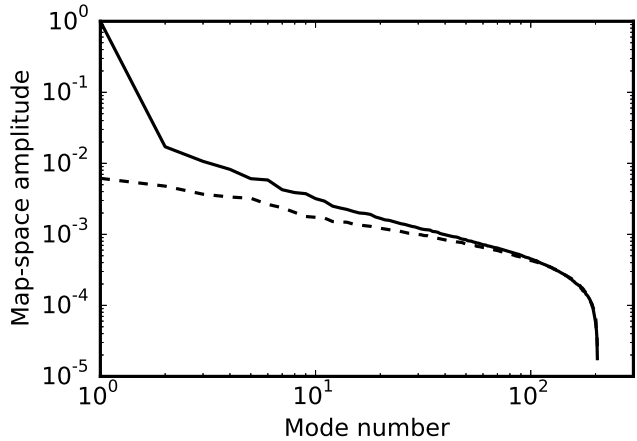}
\caption{Square root of the eigenvalue spectrum of the $\nu,\nu'$ covariance of the GBT-wide field, normalized to one for the largest foreground mode. The square root converts variance to rms temperature fluctuations in the map. The solid line shows the spectrum of the $\nu,\nu'$ covariance of the input maps, while the dashed line shows the spectrum of the $\nu,\nu'$ covariance of the difference between maps of subseasons. The difference of subseason maps removes astronomical signal and foregrounds that are common across observations and isolates any time-varying noise or instrumental systematics (Section~\ref{ssec:crosspwr}). The largest $10^3$ of the rms can be explained by $\sim 10$ modes. Beyond that peak is a plateau of much smaller modes, which are increasingly dominated by noise. See Figure~\ref{fig:eigenvectors} for the first five eigenvectors. The downturn at high mode number reflects the impact of the finite number of spectrometer channels, with some fraction cut due to RFI contamination.
\label{fig:eigenvalue_spectrum}}
\end{figure}

Figure~\ref{fig:eigenvalue_spectrum} shows the eigenvalue spectrum of foregrounds and noise in the GBT-wide survey. After removing 10 modes, the foreground fluctuations in map space are suppressed by $\sim 10^3$; however, the covariance remaining in the map is spread in a tail of modes at lower amplitude. Figure~\ref{fig:eigenvalue_spectrum} also shows the eigenvalue spectrum of the $\nu,\nu'$ covariance of the difference between maps of subseasons (Section~\ref{ssec:crosspwr}), which cancels constant astronomical signal. Before taking the map difference, we recalibrate each line of sight to isolate variation in spectral shape rather than amplitude. Beyond the brightest $\sim 10$ modes, instrumental noise and time-varying spectral response become increasingly significant. Note that the eigenvalue spectrum of the sum of covariances (noise, signal, and foreground) is generally not the sum of the spectra of their respective covariances. Hence, the the total eigenvalue spectra cannot be rigorously decomposed into the sum of signal, foreground, and noise parts. A well-designed experiment should have the majority of foreground covariance explained by a few modes. Section~\ref{ssec:aggressive} describes general conditions in which signal can be recovered well. 

\begin{figure}[htb]
\epsscale{0.5}
\includegraphics[scale=0.6]{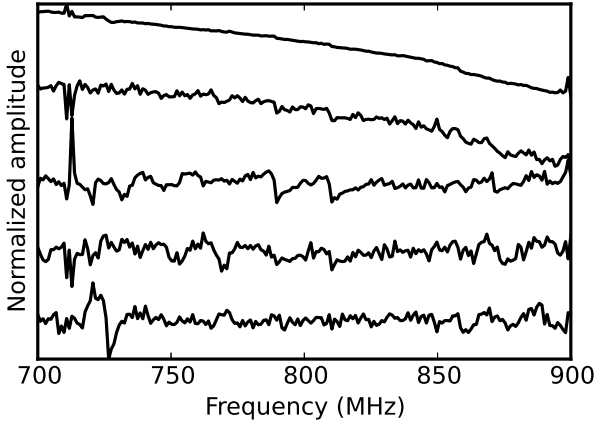}
\caption{First five eigenvectors of the $\nu,\nu'$ covariance in the GBT-wide field, descending in amplitude from the top. The largest mode is the mean synchrotron emission across the map. Despite best efforts at calibrating, there is visible small structure and non-power-law behavior. The second-largest mode is due to the chromatic beam response and can be effectively nulled by convolving the maps to a common resolution. The third mode has amplitudes that agree well spatially with polarized emission in the survey area, and the frequency structure corresponds approximately to the frequency dependence of polarization to intensity leakage. Beyond this mode, no clear instrumental response systematics can be discerned. Possibilities are calibration instability or variations in rotation measure appearing through polarization leakage. To reach the level of the signal, $\sim 30$ modes need to be removed. The spectral modes are very poorly approximated by a series of smooth polynomials. 
\label{fig:eigenvectors}}
\end{figure}

Figure~\ref{fig:eigenvectors} shows the largest five eigenvectors of the GBT-wide field. The largest mode is effectively the mean synchrotron emission across the survey. Commonly proposed smooth spectral functions (polynomials, power laws, foreground model eigenvectors) fail to explain both the small glitches and the overall, non-power-law structure. These residuals would still be hundreds of times larger than the signal. It is possible to recalibrate the data such that the largest mode is spectrally smooth, based on the largest eigenvector. However, this does not improve the overall prospects for foreground deweighting. By the construction of the eigenvectors, the higher modes are specifically orthogonal to the variations of the first. Even if the mean synchrotron were forced to be smooth, it would have no impact on the remaining, independent contaminant modes. The brightest modes are very poorly described by standard series of orthogonal polynomials. At high enough order, a complete polynomial basis could explain any of these foregrounds; however, it would also explain and project out the signal. By discovering contaminated modes in the data themselves, we remove variance surgically-leaving the most remaining degrees of freedom for signal to transfer through.

Most of the structure in Figure~\ref{fig:eigenvectors} is due to instrumental response. This section reviews the beam and spectral calibration process employed with the GBT data, emphasizing aspects relevant to contaminant modes. Complete details can be found in \citet{2013PhDT.......570M}.

A broadband noise source injects power at the feed point and acts as a stable flux reference. We switch this noise source with a rapid $64$\,ms period that allows the measured calibration signal to be uncontaminated by sky signal and RFI.  The data from a single scan (which for the wide-field data used here is 2 minutes in length) are referenced to the mean amplitude of the noise calibrator signal. The noise calibrator itself is assumed to be perfectly stable. The noise calibration (a time transfer standard) is then referenced using a collection of scans of well-characterized, bright, unpolarized point sources such as 3C\,48, 3C\,295, and 3C\,147. This procedure is performed independently in each spectral channel and for the power from the X and Y polarized antennae. Separately calibrating X and Y signals mitigates leakage of polarization into the summed unpolarized intensity. Analysis of the point source data suggests that the primary source of polarization leakage on boresight is due to a difference in gain between the X and Y antennae.

The above procedure achieves $0.5\%$ uncertainty per band over 1 minute of integration across $84$ total hours of integration in the GBT wide field survey. The data used here assume that the noise calibrator is completely stable over the observation. Similar systems \citep{2011ITMTT..59.2117B} are known to vary at the $\sim 1\%$ level. Future work must characterize the stability of noise calibrator and the covariance of its spectrum. Variations in the calibrator's spectral structure could result in a proliferation of contaminant degrees of freedom, while common mode variations can be more benign. The derived calibration in our GBT data varied at the $1\%$-level and may be partly attributable to the calibration source rather than receiver stability.

The GBT prime-focus beams have significant polarization leakage off boresight because of the off-axis design. Polarization leakage between linear polarization and total intensity is of order $5$\% of the primary gain. The polarization leakage can cause spectral structure in the foreground (and thus additional degrees of freedom) in two ways: (1) Faraday rotation of polarized synchrotron emission can vary across lines of sight and mix to frequency structure in the intensity spectrum and (2) leakage from polarization to intensity has its own spectral structure due to the instrument. The latter is conclusively observed in the third mode of Figure~\ref{fig:eigenvectors}.

Mitigation of polarization to intensity leakage is a subject for future work. A promising avenue is that the Mueller leakage beams are observed to be approximately odd functions about the boresight axis. Smoothing the maps above the beam scale tends to suppress this leakage. Spatial smoothing decreases the number of degrees of freedom that the cosmological signal can exercise. We will argue below that there is a significant penalty for fewer signal degrees of freedom because of spurious correlation of the signal and foregrounds. The most promising approach may be to pursue the ``deprojection" approach used in the BICEP CMB polarization analysis \citep{2008SPIE.7020E..1DT, 2014PhRvD..89f2006K} where contamination from leakage could be removed in the time domain data using a model of the response and polarized emission.

\subsection{Perturbative Expansion for Signal-foreground Correlations}
\label{ssec:pertsf}

Write out the cleaned map for small signals as
\begin{eqnarray}
\mat{X}_{\rm clean} &=& \mat{\Pi}^\nu_{\rm s+f}(\smap + \fgmap)  \\
&=& [1 - (\fgfmodes + \mat{\Delta}) \mat{S} (\fgfmodes + \mat{\Delta})^T ] (\smap + \fgmap) \nonumber \\
&\approx& \mat{\Pi}^\nu_{\rm f} \smap + \mat{\Pi}^\nu_{\rm f} \fgmap - \mat{\Delta} \mat{S} \fgfmodes \fgmap - \fgfmodes \mat{S} \mat{\Delta}^T \fgmap. \nonumber 
\label{eq:cleanedmap}
\end{eqnarray}
Here $\mat{\Pi}^\nu_{\rm f} \smap$ and $\mat{\Pi}^\nu_{\rm f} \fgmap$ are signal and foreground cleaned by pure foreground modes and are analogous to the direct loss in Equations~\ref{eq:xidirect} and \ref{eq:fgdirectloss}, respectively. Combine the remaining terms as
\begin{equation}
\tilde \smap \equiv -\mat{\Delta} \mat{S} \fgfmodes \fgmap - \fgfmodes \mat{S} \mat{\Delta}^T \fgmap.
\end{equation}
The combination $\mat{\Pi}^\nu_{\rm f} \fgmap + \tilde \smap$ represents residual foregrounds in the map after the empirical foreground cleaning is applied. In $\mat{\Pi}^\nu_{\rm f} \fgmap$, both $\mat{\Pi}^\nu_{\rm f}$ and $\fgmap$ are assumed to be unrelated to the signal, while the second term $\tilde \smap$ is a scaled version of the perturbations to the foreground modes due to the signal, $\mat{\Delta}$. Considering the minus signs, these residuals are anticorrelated with the signal. We will show that this term biases the band powers in a way that does not average to zero over many signal realizations, analogous to the ILC bias (e.g. \citet{2009MNRAS.397.1355E}).

When estimating the impact of empirical foreground cleaning on the signal, it is insufficient to consider only the direct loss of the signal as $\mat{\Pi}^\nu_{\rm s+f} \smap$. One must also include the impact of spurious correlations of the signal and foregrounds. These terms are only manifest if we simulate $\mat{\Pi}^\nu_{\rm s+f}(\smap + \fgmap)$, including the foregrounds.

As before, to get a rule of thumb, take the cross-correlation between the cleaned map and pure signal, as 
\begin{eqnarray}
\xi_{\rm clean} &=& N_\theta^{-1} \langle Tr(\mat{X}_{\rm clean} \smap^T) \rangle \\
&=& N_\theta^{-1} \langle Tr(\mat{\Pi}^\nu_{\rm f} \smap \smap^T) \rangle + N_\theta^{-1} \langle Tr(\tilde \smap \smap^T) \rangle, \nonumber
\end{eqnarray}
where $\langle \rangle$ is over signal realizations. We have neglected the piece $\langle Tr(\mat{\Pi}^\nu_{\rm f} \fgmap \smap^T) \rangle$ because the foregrounds cleaned with pure foreground modes $\mat{\Pi}^\nu_{\rm f} \fgmap$ have no expected correlation to the signal. Equation~\ref{eq:signallosstrace} describes the first term as being the signal and some direct loss. However, now that foreground modes are estimated from the map itself, the final term $N_\theta^{-1} \langle Tr(\tilde \smap \smap^T) \rangle$ describes the spurious average correlation of foreground residuals and the signal. Writing these out,
\begin{equation}
\xi_{\rm clean} = \xi_s + \xi_{\rm direct} + \xi_{\rm spur}.
\end{equation}
We now calculate $\xi_{\rm spur}$ in a perturbative limit and show that it significantly impacts the signal. At first order, the perturbed vectors $\mat{U}_{\rm s+f}$ are a linear combination of the pure foreground eigenvectors $\fgfmodes$ times a weight related to the perturbation. For small signals, the perturbation to the covariance is $\mat{C}_\Delta = N_\theta^{-1} (\fgmap \smap^T + \smap \fgmap^T)$, related to the spurious correlation of signal and foreground over a finite patch of the sky with $N_\theta$ samples. According to first-order perturbation theory, the vectors $\vect{\delta}_i$ in the perturbation matrix $\mat{\Delta}$ of $\mat{U}_{\rm s+f} = \fgfmodes + \mat{\Delta}$ are 
\begin{eqnarray}
\vect{\delta}_i &=& \sum_j \frac{(\vect{u}^{\rm f}_i)^T \mat{C}_\Delta \vect{u}^{\rm f}_j}{\lambda^{\rm f}_i - \lambda^{\rm f}_j} \vect{u}^{\rm f}_j. 
\end{eqnarray}
Recall that the foreground eigenvectors $\{ \vect{u}^{\rm f}_i \}$ are assumed to have full rank and are the columns of $\fgfmodes$. Through some algebra (Appendix~\ref{app:cleananticorr}),
\begin{equation}
\xi_{\rm spur} = -N_\theta^{-1} \left \langle \sum_{\substack{i\in cuts \\ j \notin cuts}} \frac{[(\vect{u}^{\rm f}_i)^T (\fgmap \smap^T + \smap \fgmap^T) \vect{u}^{\rm f}_j]^2}{\lambda^{\rm f}_i - \lambda^{\rm f}_j} \right \rangle 
\label{eqn:xispur}
\end{equation}
The terms $\fgmap \smap^T$ describe spurious correlation of signal and foreground and fluctuate about zero across signal realizations in the $\langle \rangle$ average. The fact that this expression is squared inside the $\langle \rangle$ means that there is a net anticorrelation of the foreground residuals with the signal itself. This anticorrelation depends in detail on the particular foreground $\fgmap$ in the map region. Section~\ref{ssec:fgtrans} considers this numerically for the GBT-wide survey, but it is convenient to have an analytic rule of thumb for the signal that is attenuated as a function of number of modes removed $N_m$. Under assumptions similar to Equation~\ref{eq:givenfgrule}, Appendix~\ref{app:cleananticorr} finds
\begin{equation}
\xi_{\rm clean} = \left ( 1 - \frac{N_m}{N_{\rm res, \nu}} \right ) \left ( 1-\frac{N_m}{N_{\rm res, \theta}} \right ) \xi_s,
\label{eq:emplossthumb}
\end{equation} 
where $N_{\rm res, \nu}$ and $N_{\rm res, \theta}$ are the number of modes in the frequency ($k_\parallel$) and angular ($k_\perp$) directions, respectively. The general scaling $1 - N_m/N_{\rm res, \nu}$ is intuitive: if there are only 10 spectral degrees of freedom in the signal and we need to remove 10 spectral modes, no signal remains. The dependence on the angular component is less intuitive because our cleaning operation acts entirely in the frequency direction. Indeed, if the foreground modes were taken as given, the signal loss scales as Equation~\ref{eq:givenfgrule}. The dependence on the number of angular modes $N_{\rm res, \theta}$ arises as a by-product of measuring the foreground spectral modes from the map itself. The covariance $\mat{\hat C} = N_\theta \mat{X} \mat{X}^T$ measures foreground spectral modes against a limited number of signal realizations. In general, if the same foreground mode is observed against many different signal realizations, the spurious correlation averages down.

Equation~\ref{eq:emplossthumb} is only a rule of thumb, and simulations are needed to effectively measure the number of resolution elements $N_{\rm res, \nu}$ and $N_{\rm res, \theta}$. These quantities are not simply $N_\nu$ and $N_\theta$, the number of spectral and angular pixels in the survey. Instead, $N_{\rm res, \nu}$ and $N_{\rm res, \theta}$ roughly relate to the number of modes in the signal at the $\vect{k}$ scales of interest. Signal components at the lowest $k_\perp$ of the survey have large wavelengths and few $N_{\rm res, \theta}$ modes in the survey volume. The spurious correlation of signal and foreground does not average down over many signal modes, and so most of the large-scale signal is lost. The impact of foreground cleaning on the signal depends strongly on both $k_\perp$ and $k_\parallel$, and so must be treated as a 2D transfer function.

From the point of view of survey design, one wants to maximize the number of resolution elements $N_{\rm res, \nu}$ and $N_{\rm res, \theta}$. The beam size fixes the ultimate sensitivity at high $k_\perp$, and having high $N_{\rm res, \theta}$ translates into large surveys with many beam spots. In general, intensity mapping surveys will need to cover larger areas than Fisher estimates from thermal noise suggest, owing to the fact that potentially many degrees of freedom in the data need to be used to estimate foregrounds.

\subsection{Estimating the Signal Loss Transfer Function}
\label{ssec:fgtrans}

The previous section argued that when contaminated modes are discovered from the data themselves, residual foregrounds will be anticorrelated with the signal (on average). The means that to assess signal attenuation due to foreground cleaning, it is insufficient to apply the cleaning to signal-only simulations and measure loss as in Section~\ref{sec:beamimpact}. Instead, the foreground cleaning should be applied to simulations of signal plus foregrounds. A reasonable approach is to add signal simulations $\vect{x}_{\rm sim}$ to the measured map itself, $\vect{x}$ (which is assumed to be dominated by continuum foregrounds). One can then subtract the foreground power back from the band power, as
\begin{eqnarray}
\hat q_\alpha |_{\rm out} &=& [\mat{C}^{-1} \mat{\Pi}^\nu_{\rm s+f}(\vect{x} + \vect{x}_{\rm sim})]^T \mat{Q}_\alpha \mat{C}^{-1} \mat{\Pi}^\nu_{\rm s+f}(\vect{x} + \vect{x}_{\rm sim}) \nonumber \\
&&- (\mat{C}^{-1} \mat{\Pi}^\nu_{f} \vect{x})^T \mat{Q}_\alpha \mat{C}^{-1} \mat{\Pi}^\nu_{f}\vect{x},
\label{eq:noisyfgloss}
\end{eqnarray}
where the modes of the cleaning operation $\mat{\Pi}^\nu_{\rm s+f}$ are determined from the map of real data plus simulations ($\vect{x} + \vect{x}_{\rm sim}$) and $\mat{\Pi}^\nu_{f}$ is determined from only the real data, taken to define the foreground modes. Equation~\ref{eq:noisyfgloss} represents the band power estimate of the remaining signal simulation, and the transfer function can be estimated as the ratio of this quantity to the input simulation band power weighted by the thermal noise of the survey, or 
\begin{equation}
\hat q_\alpha |_{\rm in} = (\mat{C}^{-1} \vect{x}_{\rm sim})^T \mat{Q}_\alpha \mat{C}^{-1} \vect{x}_{\rm sim},~~~~~~T_\alpha^F = \left \langle \frac{\hat q_\alpha |_{\rm out}}{\hat q_\alpha |_{\rm in}} \right \rangle.
\label{eq:basicfgloss}
\end{equation}
This is analogous to Equation~\ref{eq:fgtrans} but includes spurious correlation of signal and foreground. Cosmological signals with poorly understood amplitudes (such as when a cross-power with a galaxy survey is not available) may need a range of signal simulation amplitudes to understand any sensitivity of the transfer function.

The optimal number of modes to remove is described next in Section~\ref{ssec:aggressive}. To be able to compare outcomes for different numbers of modes removed, the transfer function needs to be calculated for a range of scenarios. Especially for too few modes removed, the residual foreground variance may be significant. In this case,  the estimate of $\hat q_\alpha |_{\rm out}$ is noisy and Equation~\ref{eq:basicfgloss} requires an average over many realizations to converge satisfactorily. 

Write out the cleaned map of Equation~\ref{eq:cleanedmap} in $\vect{x}$ rather than $\mat{X}$ notation as 
\begin{equation}
\vect{x}_{\rm clean} = \mat{\Pi}_{\rm s+f}(\vect{x} + \vect{x}_{\rm sim}) \approx \mat{\Pi}_{f} \vect{x} + \mat{\Pi}_{f} \vect{x}_{\rm sim} + \vect{\tilde x}_{\rm sim},
\end{equation}
where again $\vect{\tilde x}_{\rm sim}$ is the critical piece of residual foregrounds that are anticorrelated with simulated signal. On average, the term $\mat{\Pi}_{f} \vect{x}$ is just producing variance and contains nothing related to the signal, so it can be subtracted. We can then form $\mat{\Pi}_{\rm s+f}(\vect{x} + \vect{x}_{\rm sim}) - \mat{\Pi}_{f} \vect{x} = \mat{\Pi}_{f} \vect{x}_{\rm sim} + \vect{\tilde x}_{\rm sim}$ to get just those pieces relevant to understanding how the signal acts under foreground subtraction. Signal loss can then be assessed in the cross-power between this cleaned signal map and the input signal
\begin{equation}
\hat q_\alpha |_{loss} = [\mat{\Pi}_{\rm s+f}(\vect{x} + \vect{x}_{\rm sim}) - \mat{\Pi}_{f} \vect{x}]^T \mat{C}^{-1} \mat{Q}_\alpha \vect{x}_{\rm sim} \label{eq:qloss}.
\end{equation} 
We refer to this as a one-sided estimate of signal loss, because $\mat{\Pi}$ is only on one side of the quadratic estimator. The numerator of the transfer function needs the expectation value of signal with foreground cleaning applied to both sides. We can rewrite this using a relation between trace and covariance as 
\begin{eqnarray}
&&Tr(\mat{C}_{,\alpha} \mat{\Pi}^T \mat{C}^{-1} \mat{C}_{,\alpha} \mat{C}^{-1} \mat{\Pi}) \\ \nonumber &=& \frac{1}{2} {\rm Cov} (\vect{x}^T \mat{\Pi}^T \mat{C}^{-1} \mat{C}_{,\alpha} \vect{x}, \vect{x}^T \mat{\Pi}^T \mat{C}^{-1} \mat{C}_{,\alpha} \vect{x}) 
\end{eqnarray}
where $\vect{x}$ is normally distributed with covariance $\mat{1}$. For Gaussian $\vect{x}$, the covariance can be approximated as the $2 \nu(k)^{-1} P^2_\alpha \delta_{\alpha, \beta}$, where $P_\alpha$ is the mean power spectrum (in this case $\langle \vect{x}^T \mat{\Pi}^T \mat{C}^{-1} \mat{C}_{,\alpha} \vect{x} \rangle$) and $\nu(k)$ is the number of modes in the survey volume. A similar expression holds for the denominator of the transfer function. The numerator can be estimated in Monte Carlo with lower noise by removing the residual foreground variance as in Equation~\ref{eq:qloss}. The $\nu(k)$ pre-factors drop out in the ratio, and the transfer function can then be expressed as the average of the ratios
\begin{equation}
T^F_\alpha \approx \left \langle \frac{
[\mat{\Pi}_{\rm s+f}(\vect{x} + \vect{x}_{\rm sim}) - \mat{\Pi}_{f} \vect{x}]^T \mat{C}^{-1} \mat{Q}_\alpha \vect{x}_{\rm sim}}{\vect{x}_{\rm sim}^T \mat{C}^{-1} \mat{Q}_\alpha \vect{x}_{\rm sim}} \right \rangle^2.
\label{eqn:fgtranssim}
\end{equation}

Equation~\ref{eqn:fgtranssim} provides a convenient procedure for estimating signal attenuation due to cleaning empirically determined spectral modes. 

\begin{figure}[htb]
\epsscale{0.5}
\includegraphics[scale=0.6]{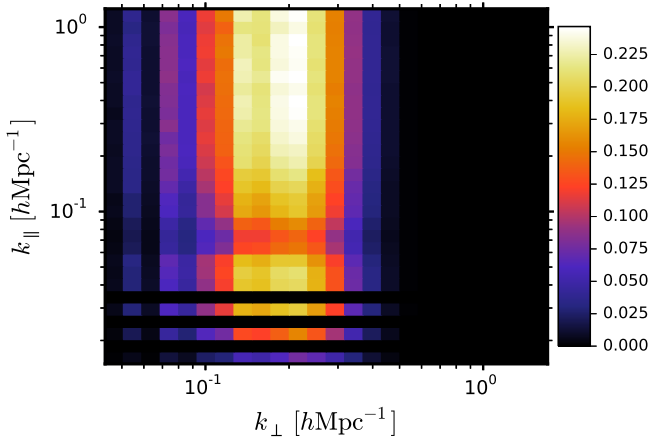}
\caption{Transfer function of signal attenuation for 30 foreground modes removed in the GBT-wide field, including the effect of the beam. When contaminated modes are determined from the data themselves, it is essential to include foregrounds in simulations of the signal attenuation. This can be viewed two ways: either (1) the signal perturbs the foreground modes or (2) spurious correlations between signal and foreground need to be included. Especially on $k$-modes that approach the size of the survey region, there are few modes available and most of the signal is lost. To reach the lowest upper bound of the signal auto-power, foreground cleaning also removed much of the signal. The beam rolls off response at high $k_\perp$, and the signal attenuation at low $k_\perp$ and $k_\parallel$ is due to the spectral foreground removal.\label{fig:fg_transfer}}
\end{figure}

Figure~\ref{fig:fg_transfer} shows the transfer function for the GBT-wide data (including the beam). At high $k_\perp$, signal is lost to large beam size. The signal attenuation at low $k_\perp$ and low $k_\parallel$ is due to spectral foreground cleaning. Signal attenuation toward low $k_\parallel$ can be explained as ``direct" loss of removing functions along the line of sight that have overlap with the signal's spectral variation (Section~\ref{ssec:directloss}). The structure at $k_\parallel = 0.07~h{\rm Mpc}^{-1}$ originates from additional foreground degrees of freedom around that scale that needed to be nulled. The signal loss at low $k_\perp$ is less intuitive and is due to the spurious spatial correlation of signal and foreground on large angular scales. Each foreground spectral mode $\vect{u}^{\rm f}_i$ discovered in the survey is associated with a spatial mode through the singular value decomposition (SVD) $\mat{X} = \sqrt{N_\theta} \mat{U} \mat{\Lambda}^{1/2} \mat{V}^T$ (see Appendix~\ref{app:cleananticorr} and \citet{Nityananda10}). The spectral functions nulled in the line of sight are associated with spatial modes of the signal. The largest foregrounds are spatially smooth and share some spurious correlation with the small number of spatial signals at low $k_\perp$. Nulling the spectral variation associated with these spatial modes erases the large spatial scales. Signal loss toward low $k_\perp$ is an artifact of nulling spectral modes determined from the map itself.

\subsection{Aggressiveness of Foreground Cleaning}
\label{ssec:aggressive}

\begin{figure*}[htb]
\epsscale{0.5}
\includegraphics[scale=0.6]{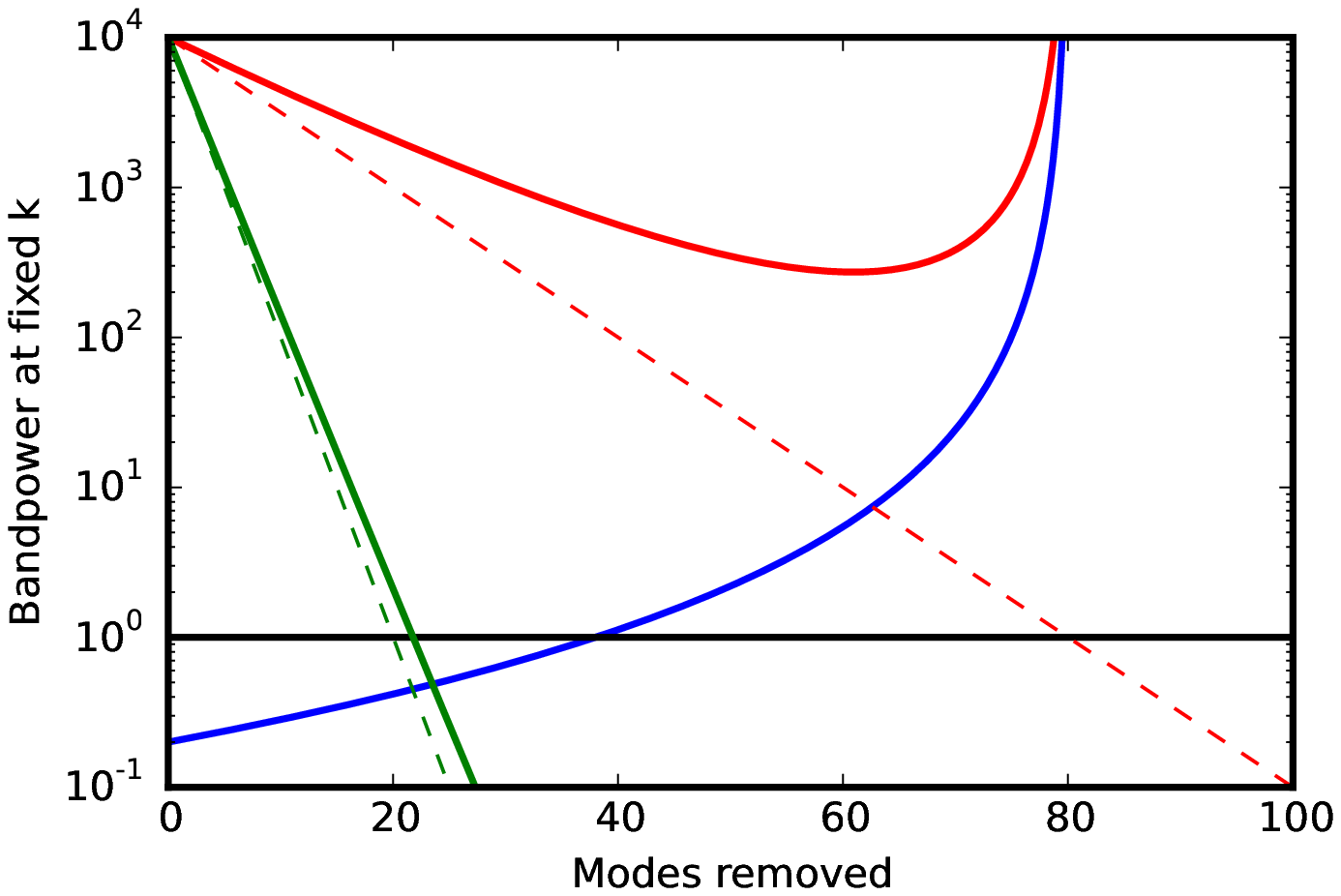}
\includegraphics[scale=0.6]{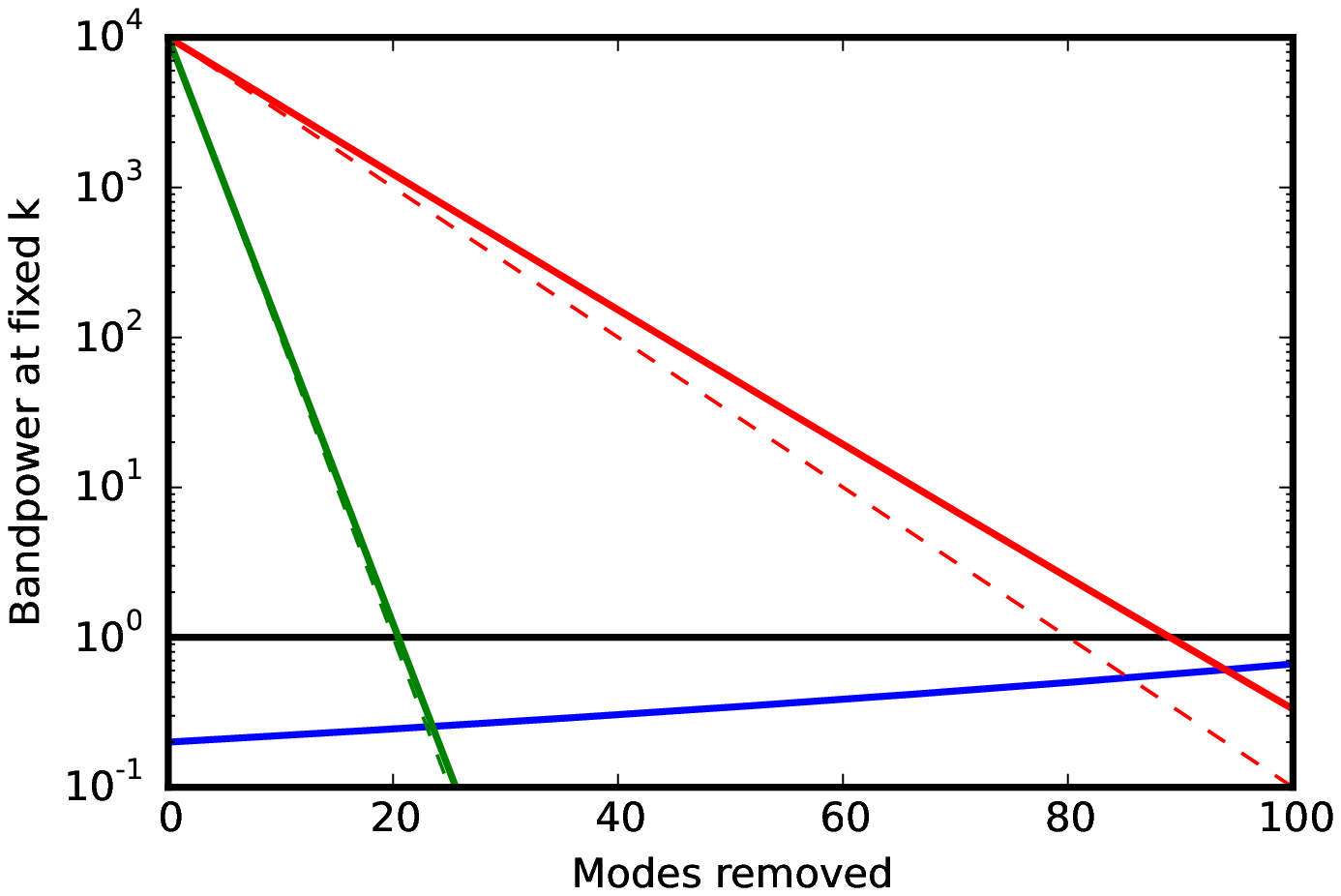}
\caption{Intensity mapping foreground regimes. The signal power at a fixed $k$ band power is a constant (black) normalized to 1. The eigenvalue spectrum of foregrounds is approximated as a power law, with a steep (green) and shallow (red) decline from $10^4$ times the signal (in power, not map space), as more modes are removed. Errors on the band power (blue) increase as more signal is removed by foreground cleaning, according to Equation~\ref{eq:emplossthumb}. The noise level is set to achieve a $5\sigma$ detection of the signal at zero modes removed.  Residual foregrounds are also boosted by the signal loss transfer function. Dashed red and green show the decline of the intrinsic foreground variance as more modes are removed. The solid lines show the residual foreground variance with transfer function applied to keep signal constant. A successful experiment will have foregrounds below noise and noise below signal. The left panel shows an experiment that has 80 spatial degrees of freedom for this $k$ mode. This experiment will only be successful if the foreground eigenvalue spectrum falls rapidly (green). Note that for the shallow spectrum, the contamination at its minimum is hundreds of times the signal and larger than the noise. The right panel shows an experiment that has $250$ degrees of freedom for the $k$ mode. In this case, there are sufficient modes to secure a detection even for the shallow foreground eigenvalue spectrum. This does not include the impact of noise on contaminated mode determination, which generally limits the maximum number of modes that can be removed, or alternately how far the foreground can be suppressed relative to thermal noise.
\label{fig:im_plausibility}}
\end{figure*}

For residual foregrounds to be negligible, (1) the foreground must inherently have few spectral degrees of freedom compared to the signal and (2) instrument response must be well constrained and mix bright foregrounds into a limited number of new modes. Some instrument responses like calibration stability will inevitably produce a long tail of new modes where each line of sight observes a slightly different continuum. This produces high-rank residuals that cannot be easily estimated or subtracted.

Until proven otherwise, interpretation of intensity mapping power spectra should account for the additive bias from residual/unweighted foregrounds from misspecification or incomplete knowledge of the foreground covariance. The goal of good instrument design and calibration is to push the amplitude of residual modes well below the thermal noise in the maps, so that the biases are at the level of the power spectral errors. (For some goals with particular $P(k)$ shapes like the baryon acoustic oscillation, requirements on the additive bias may be weaker.)

Cross-correlation with a spectroscopic galaxy catalog has the advantage that residual foregrounds will boost errors rather than producing an additive bias. Galaxy densities and line intensity fields will not be perfectly correlated, so cross-correlation provides a lower bound on the fluctuation power in the line survey. Combined with the auto-power of the line survey (which is an upper limit due to its additive bias), an intensity survey can provide an indirect inference on the range of true fluctuation power in the atomic line \citep{2013MNRAS.434L..46S}.
 
Foreground cleaning requirements vary between auto-power surveys of intensity maps and cross-powers with side surveys. For too few modes removed, the cross-power errors are large because of the map variance from foregrounds. By removing more than the optimal number of modes, the error bars increase again as more signal is lost \citep{2013ApJ...763L..20M}. The optimal cleaning is achieved for the lowest errors on the cross-power, a well-defined quantity. Similarly, the additive bias on the auto-power of the intensity mapping survey drops as more modes are removed, but errors again increase. The upper bound on the line fluctuation intensity is the band power plus the error. One can choose a number of modes to remove that gives the lowest upper bound on the auto-power amplitude. Generally, the foreground cleaning requirements for the cross-power with a side survey are much less stringent than the auto-power of the intensity survey. Each Fourier mode of the cross-power has a different realization of the residual foregrounds crossed with the galaxy survey. In a survey with many $k$-modes contributing to a band power, the cross-powers can average down over these correlations to produce small errors. In contrast, each mode of the auto-power of the intensity map with itself is a sample of approximately the same power, which does not average down.

Figure~\ref{fig:im_plausibility} shows a high-level summary of the plausibility of detecting the auto-power in line intensity surveys. It is clear that one wants (1) a survey that has many signal degrees of freedom in the $k$-modes of interest and (2) a shallow eigenvalue spectrum of the foregrounds. These two goals can be contradictory. The first goal suggests coverage of wide areas of the sky to suppress spurious correlations between signal and foreground. However, in surveying large areas, an instrument may observe a wider range of foreground spectral modes or be prone to instrumental effects that are harder to control across large areas and time. These effects generally boost the number of modes that need to be estimated. Possible examples of these effects are spatial variations in spectral index, variations in bandpass calibration, or exposure to a wider range of rotation measures through polarization to intensity leakage. 

CMB B-modes have had a vigorous history of studies of instrumental systematics (e.g. \citet{2003PhRvD..67d3004H}). Similar studies should be undertaken for intensity mapping surveys. These would characterize the impact of calibration stability, beam response, or other instrumental response to foregrounds. This requires detailed simulations that are beyond the scope of this methods paper. Foreground emission and variations in spectral index may contribute a handful of spectral degrees of freedom, but instrumental effects have the potential to mix these into many more bright degrees of freedom that are not well known in advance. In future studies, the eigenvalue spectrum of input astronomical foregrounds can be compared to the eigenvalue spectrum as observed by a simulated instrument. A well-designed survey will control the number of modes induced by the instrument response, targeting the rapidly falling green contours in Figure~\ref{fig:im_plausibility}.

\section{Assembling the Final Product}
\label{sec:finalprod}

Previous sections defined the core aspects of the estimator and foreground cleaning. In practice, the estimator can be made more robust to temporally variable noise by forming cross-powers between subseasons. The 2D band powers derived above are expected to have low signal-to-noise ratio in the first generation of experiments. They are also are difficult to interpret. We describe a procedure to optimally bin onto 1D powers and develop a covariance model and decorrelation of those powers for convenient display. The binning weights from 2D to 1D provide insight into the information content of the intensity survey.

\subsection{The Subseason Cross-power}
\label{ssec:crosspwr}

Thermal noise is uncorrelated between subseasons of the observations to an excellent approximation. In addition, some forms of contamination such as time-varying RFI, or calibration instability (and its induced foreground residuals), may be largely uncorrelated across times. The cross-powers between $N_s$ split subseasons of the data $\vect{x}_A ... \vect{x}_{N_s}$ therefore have no additive noise bias from these terms \citep{2005MNRAS.358..833T, 2013MNRAS.434L..46S, 2014PhRvD..89b3002D}. A subseason cross-power estimator is formally nonoptimal, but does follow from the form of the optimal quadratic estimator. Appendix~\ref{app:crosspower} shows the choices that lead to the estimator
\begin{equation}
\hat q_\alpha^{A\times B} \propto (\mat{N}_A^{-1} \vect{x}_A)^T \mat{C}_{,\alpha} (\mat{N}_B^{-1} \vect{x}_B),
\end{equation}
where $\mat{N}_A$ and $\mat{N}_B$ are the nonsignal noise covariance in two maps. The data are weighted by their respective covariances for the two subseasons. Following the form of Equation~\ref{eq:avoidmode} to avoid contaminated modes,  
\begin{equation}
\hat q_\alpha^{A\times B} \propto (\mat{W}_A \mat{\Pi}_A \vect{x}_A)^T \mat{C}_{,\alpha} (\mat{W}_B \mat{\Pi}_B\vect{x}_B).
\end{equation}
This expression continues to assume that after projecting the contaminated modes through $\mat{\Pi}_A \vect{x}_A$, the remaining covariance is dominated by thermal noise, which simply follows the coverage map of the experiment in the two subseasons and can be implemented as diagonal inverse covariance weights $\mat{W}_A$ and $\mat{W}_B$. Take the projections to remove spectral modes along each line of sight as in Equation~\ref{eq:simplelosclean},
\begin{equation}
\mat{\Pi}_A \vect{x}_A = vec(\mat{\Pi}^\nu_A \mat{X}_A) = vec((\mat{1} - \mat{U}_{A, {\rm f}} \mat{S} \mat{U}_{A, {\rm f}}^T),\mat{X}_A)
\end{equation}
where now the contaminated modes $\mat{U}_{A,{\rm f}}$ can be particular to the subseason. (If the instrument noise varies with frequency, that weight should appear here as in Equation~\ref{eq:weightedlos}. We neglect it to keep a simpler equation.)

We can also develop foreground modes that are best tuned to the respective data splits by finding modes of the cross-covariance of the two subseason maps as
\begin{equation}
\mat{\hat C}_{\nu, \nu'} = \mat{X}_A \mat{X}_B^T = \mat{U}_{A, {\rm f}} \mat{\Lambda} \mat{U}_{B, {\rm f}}^T, 
\end{equation}
where the final operation is the SVD of the cross-variance. The covariance estimate can be improved by using the weighted maps $\mat{W}_A \vect{x}_A$ \citep{2013MNRAS.434L..46S}. However, in doing this it is important to force $\mat{W}_A$ to be a separable expression as $\mat{W}^\nu_A \otimes \mat{W}^\theta_A$ by averaging over the frequency and spatial directions. Otherwise, the weighting operation would increase the rank (e.g. nonseparable $\mat{W}$ applied to $\mat{X} = \vect{u} \vect{v}^T$ is no longer generally rank-1). 

Most instruments have a chromatic beam. Without accounting for this, the modal structure will include variance introduced by this frequency response. In GBT, this was the second-largest mode (Figure~\ref{fig:eigenvectors}) and was treated by convolving the maps to a common resolution. In general, any well-modeled aspects of the instrument should be treated in the map to reduce the burden of discovering those modes in the data.

The estimate $\hat q_\alpha$ is the average over all cross-power pairs 
\begin{equation}
\hat q_\alpha = \frac{1}{N_s (N_s-1)} \sum_{i,j, i\neq j} \hat q^{i\times j}_\alpha.
\label{eqn:crossed_average}
\end{equation}
We have assumed that all pairs have similar statistical power. If thermal noise dominates, this can be arranged by forming all of the subseason split maps from approximately the same integration times and areas. If this is not possible, then the cross-powers should be more optimally weighted.

\subsection{Projecting to 1D Powers}

The optimal estimator structure developed in Section~\ref{ssec:optestproc} binned from 3D Fourier space to the 2D band powers assuming that the noise was isotropic across constant $k_\perp = \sqrt{k_x^2 + k_y^2}$ under the action of $\mat{C}^{-1}$. For the first generation of intensity mapping experiments, there is likely to be insufficient signal-to-noise ratio on each 2D band power. Binned 1D power $\hat P(k)$ for a band of $\sqrt{k_\perp^2 + k_\parallel^2} \in (k, k+\Delta k)$ contains most of the cosmological information other than redshift-space distortions and so provides a convenient summary.

\begin{figure*}
\epsscale{0.5}
\includegraphics[scale=0.6]{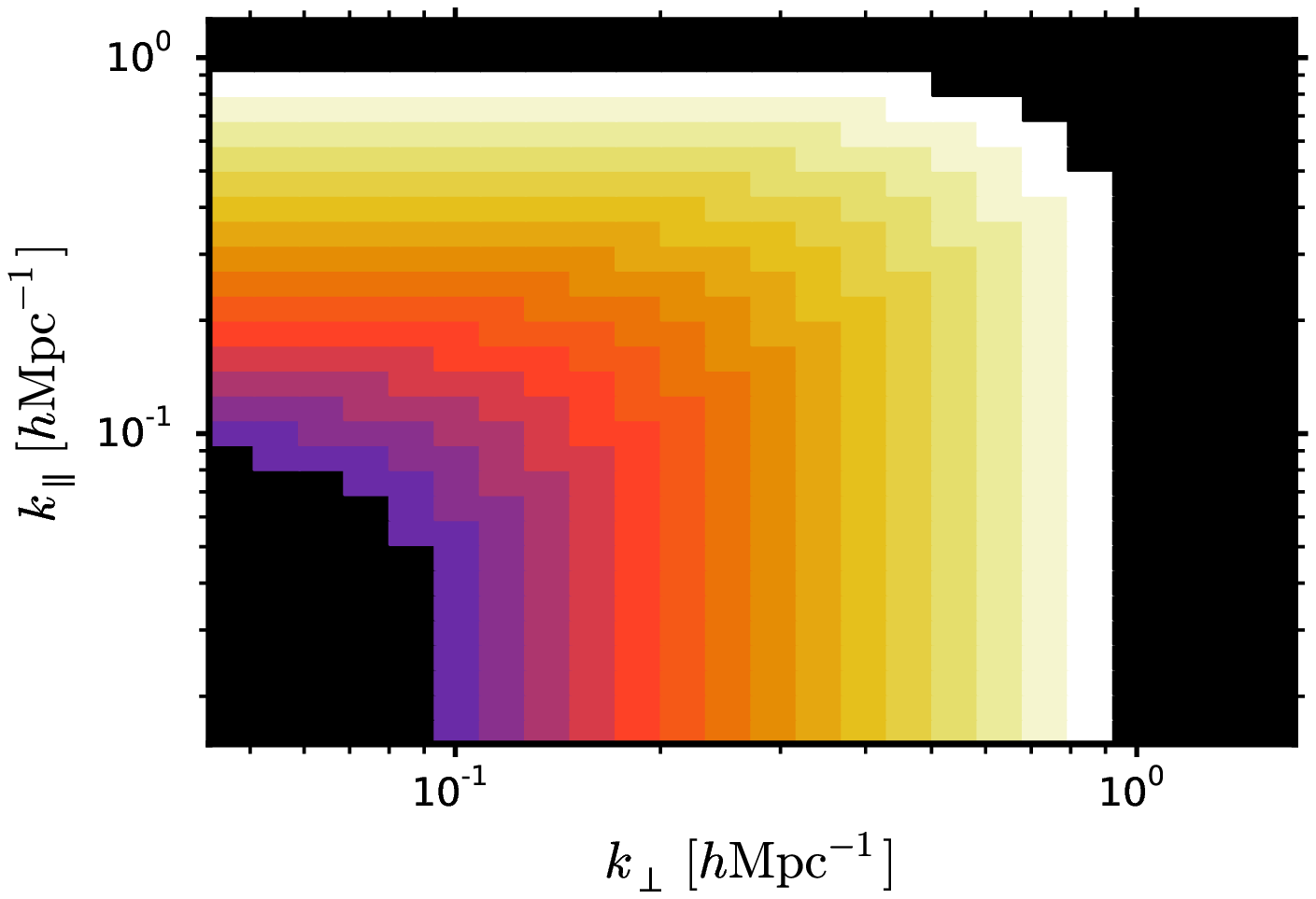}
\includegraphics[scale=0.6]{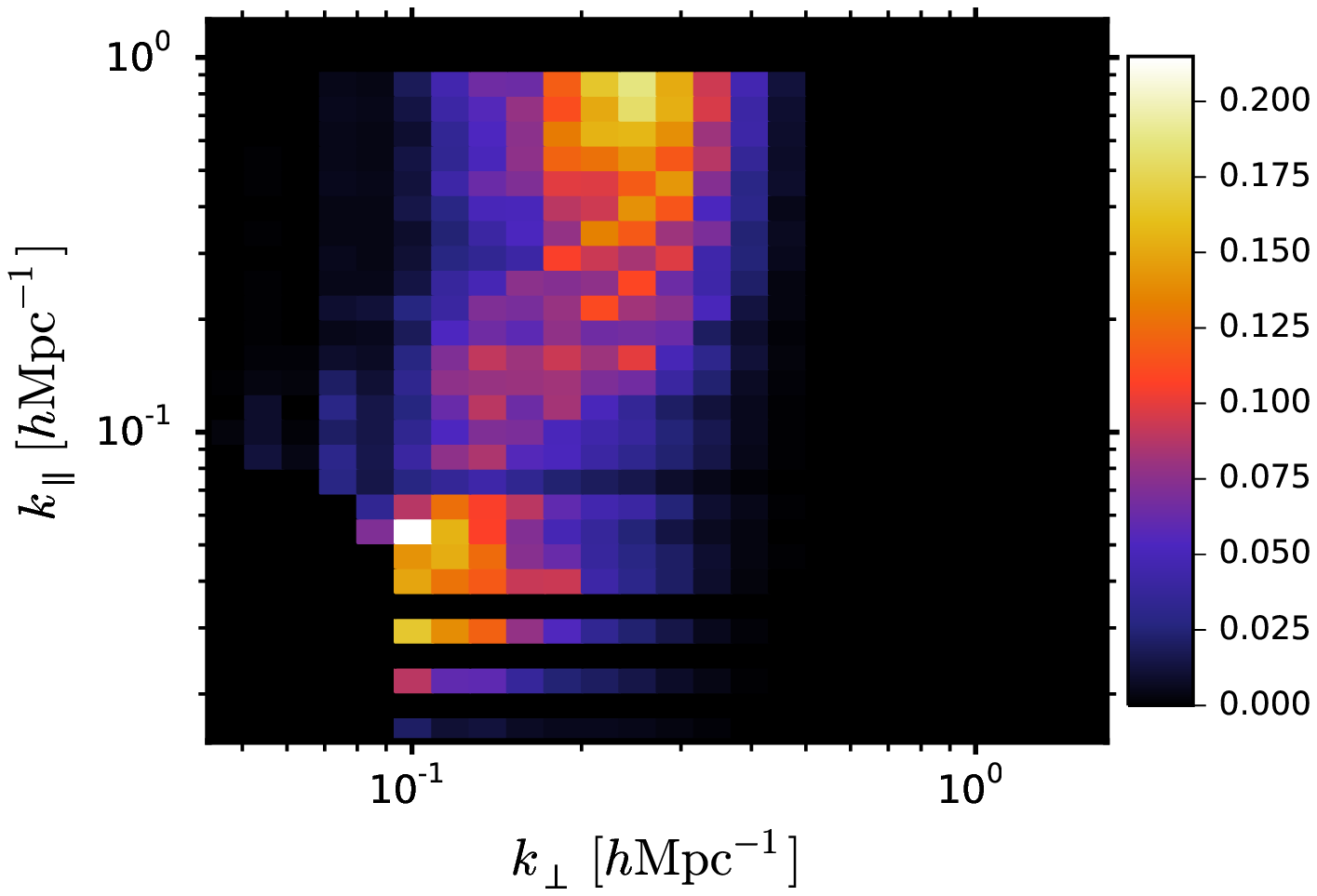}
\caption{Left: binning from 2D to 1D powers occurs in bands along constant $k = \sqrt{k_\perp^2 + k_\parallel^2}$ for 15 bands between $k=0.1~h{\rm Mpc}^{-1}$ and $1~h{\rm Mpc}^{-1}$. Constant colors show the 2D $k$-cells contributing to a 1D band power. The bins are constant in $\log(k)$. Right: the noise weight from Equation~\ref{eq:2dnoiseweight} for the GBT-wide survey. The weight bins per 2D bandpower are normalized to sum to $1$ across the 1D $k$-bands in the left panel. The colored bands in the left panel guide the eye for 2D $k$ bins that contribute. At low $k\approx 0.1~h{\rm Mpc}^{-1}$, the weight favors slowly varying spectral modes and more rapidly varying spatial modes (the vertical part of the iso-$k$ contour). This reflects the fact that the foreground cleaning destroys spatial modes with $k_\perp < 0.1~h{\rm Mpc}^{-1}$ because the bright foreground modes are associated with spatially smooth structure in the map. In contrast, long-wave spectral modes are less penalized because the contaminated modes have high-frequency spectral structure (only the largest mean synchrotron mode is spectrally smooth). In contrast, by $k \approx 1~h{\rm Mpc}^{-1}$ almost all information comes from rapidly varying spectral modes (the horizontal part of the iso-$k$ contour) because the beam has destroyed essentially all of the information at $k_\perp > 0.4~h{\rm Mpc}^{-1}$. For intermediate scales, information at a wider range of modest $k_\perp$ and $k_\parallel$ is weighted more heavily. The noisy structure of the weights is driven by variations in the auto-power in the denominator of Equation~\ref{eq:2dnoiseweight}.
\label{fig:2dweights}}
\end{figure*}

Noise is strongly anisotropic across the $k_\perp$ and $k_\parallel$ ring that contributes to some $k$. This is because the number of modes increases quadratically in $k_\perp$ and linearly in $k_\parallel$, the transfer functions will vary across the ring in 2D $k$-space, and the noise itself may vary across the ring. These factors also make it difficult to interpret the 2D power spectrum. In going from 2D to 1D powers, we can apply an inverse-covariance weight to maximize the 1D signal-to-noise ratio. Let $\alpha$ continue to refer to 2D band powers $k_{\perp, \alpha}, k_{\parallel, \alpha}$ and $R_k$ be the ring of $k_\alpha$ values that contribute to the $k$ band power. The weighted sum is
\begin{equation}
\hat P(k) = \frac{\sum_\alpha I_{\alpha \in R_k} W_\alpha \hat p_\alpha}{\sum_\alpha I_{\alpha \in R_k} W_\alpha}.
\end{equation}
Figure~\ref{fig:2dweights} shows the indicator $I_{\alpha \in R_k}$ as a set of colored bands for each $k_\alpha$ bin. Again the choice of $W_\alpha$ is a weighting related to optimality rather than bias so long as $W_\alpha$ is based on the variance rather than the mean value of the bins. In contrast, an analysis should not use $W_\alpha$ that does additional foreground suppression by masking areas of the 2D power that have apparent residual foregrounds in the mean band power, as this would produce bias. 

In the language of subseason data splits, $\hat p^{A \times A}_\alpha$ is an auto-power, and it describes the variance of thermal noise, cosmological signal, and residual foregrounds. In practice for GBT-wide, this was dominated per $k_\alpha$ bandpower cell by thermal noise. Assuming Gaussian errors from the dominant thermal noise, the inverse covariance weight per pixel is
\begin{equation}
W_\alpha = \frac{N_\alpha T_\alpha^2}{\hat p_{\rm auto, \alpha}^2},
\label{eq:2dnoiseweight}
\end{equation}
where $N_\alpha$ is the number of 3D Fourier cells contributing to the 2D power in band power $\alpha$, $T_\alpha^2$ is the total transfer function that applies to the band power $\alpha$, and $\hat p_{\rm auto, \alpha}$ is estimated from the 2D auto-powers. Figure~\ref{fig:2dweights} shows $W_\alpha$ for the GBT-wide survey. Prior to these measurements, the theoretical expectation was that foreground spectra vary slowly and that primarily information at low $k_\parallel$ is lost. In the case of foreground modes determined from the data themselves, large angular scales have more spurious correlation with foregrounds, and considerable $k_\perp$ is also lost, resulting in deweighting of low $k_\perp$. Interband correlations are estimated in the next section using simulations, but are not exploited in the 2D to 1D weighting.

\subsection{Errors and Decorrelation}
\label{eq:errmodel}

A simple estimate of the band power errors can be derived from the variance of cross-power pairs of subsets of the data. In the case of a survey split into maps $A$, $B$, $C$, and $D$, the cross-powers $A\times B$, $A\times C$, $A\times D$, $B\times C$, $B\times D$, $C\times D$ form six unique samples of the instrument's thermal noise. The subseason maps $A$, $B$, $C$ and $D$ all share the same underlying signal and residual foregrounds, so the variance of the crossed pairs does not reflect the sample variance. It is also a poor estimate of the errors if there are a limited number of crossed pairs (six in the example here).

If the cleaned maps are dominated by Gaussian fluctuations such as from thermal noise, the full bandpower covariance can be determined from the measured auto-power spectrum between common sections (e.g. $A\times A$), the power spectrum across sections, and the survey geometry (see, e.g. \citet{2011ApJ...729...62D} for an application in CMB analysis). The validity of Gaussian errors depends on both the signal and residual foregrounds. If a survey resolves sufficient non-Gaussian cosmological signal, then methods such as \citet{2012MNRAS.423.2288H} should be considered to properly describe the errors. To date, no intensity maps have reached this high cosmological signal-to-noise regime. 
Appendix~\ref{app:gauss_error} calculates the full band power covariance ${\rm cov} (\hat P, \hat P)$ using a hybrid of Monte Carlo simulations for the off-diagonal structure and Gaussian errors for the amplitudes.

With the optimal estimator, both the final covariance and the window function are the Fisher matrix. Standard discussion of decorrelation \citep{2000MNRAS.312..285H} for optimal estimators freely moves between undoing the effect of the window function through $\mat{F}^{-1}$ and diagonalizing the final band power covariance through $\mat{F}^{-1/2}$. For the suboptimal estimators here, the variance $2 Tr(\mat{C} \mat{Q}_\alpha \mat{C} \mat{Q}_\beta)$ is different from the windowing matrix $Tr(\mat{C}_{,\beta} \mat{Q}_\alpha)$. The band powers $\hat p_\alpha$ so far have just been scalar normalizations times the pseudo-powers $\hat q_\alpha$ rather than a linear combination that decorrelates the band powers.

With the full band power covariance model in hand from Appendix~\ref{app:gauss_error}, we can repeat the classic decorrelation choice \citep{2000MNRAS.312..285H} of $\mat{F}^{-1/2}$ by taking ${\rm cov} (\hat P, \hat P)_{\rm data}^{-1/2}$ to multiply by the band powers (normalized so that the weights on the band powers sum to 1). So long as decorrelation multiplies by an invertible matrix, no information is lost and the choice is purely one of display, which generally benefits from uncorrelated errors. \citet{2013MNRAS.434L..46S} and \citet{2013ApJ...763L..20M} use this pipeline for two GBT intensity mapping surveys and describe results and the interpretation of 1D band powers subject to additive bias.

\section{Discussion}
\label{sec:disc}

Intensity mapping experiments have the potential to map cosmological volumes with resolution and sensitivity requirements that are modest compared to direct spectroscopic surveys of objects. In addition to atomic or molecular line radiation, these surveys generally receive continuum radiation that can be orders of magnitude brighter. We have developed a quadratic estimator that combines some aspects of both galaxy and CMB surveys, but also accommodates methods of down-weighting bright continuum emission. A fully optimal estimator requires a model of the covariance of contamination, which we argue is not well known prior to an experiment. In the example of GBT data, the spectral structure of contaminant modes was related primarily to the instrument response rather than intrinsic spectra. The instrumental response is residual in the sense that considerable new effort was put into calibration, in addition to the heritage of a well-established instrument.

We develop an estimator that (1) takes cross-powers of subseasons to avoid bias from temporally-variable noise (2) estimates a foreground covariance model from the data itself through a reduction in dimensionality, (3) accounts for the impact of spurious correlations between signal and foreground, and (4) derives the final 1D power and its errors. Transfer functions provide a convenient way to calibrate the estimator's output to the signal input and can be estimated efficiently using Monte Carlo simulations. Spurious correlations of signal and foreground result in an average anticorrelation of signal and residual foregrounds. Simulations for the transfer function must include foregrounds to properly account for this effect. 

Figure~\ref{fig:im_plausibility} argues that for an intensity mapping experiment to be successful, it must control the eigenvalue spectrum of foregrounds and observe a large enough area that spurious correlation between signal and foreground can average down. While the intrinsic foregrounds may only have a handful of degrees of freedom, variations in instrumental response have the potential to mix those spectral modes into a larger number of new modes and a shallower eigenvalue spectrum. In particular, variable spectral calibration contributes some level of full-rank covariance (each line of sight responds differently to bright emission), even with rank-1 input contamination. The eigenvalue spectrum of the $\nu,\nu'$ covariance of the maps is the central metric for the quality of the calibration or mapping procedure. Because of instrumental effects, contaminant modes are not necessarily smooth and so generally poorly described by smooth functions. The salient aspect here is not the spectral smoothness but rather that the signal can fluctuate in many more ways than the instrument's response to bright foregrounds.

Figure~\ref{fig:im_plausibility} also demonstrates one of the challenges of reaching convincing detection using intensity mapping data alone, when no cross-power corroboration is possible. An experiment only has access to the band power estimate of signal plus foreground, which formally represents an upper bound on signal. This total bandpower falls as foregrounds are more aggressively cleaned. If signal dominates, the bandpower will reach a plateau where errors increase, but the amplitude does not diminish as the cleaning pushes to down-weight more foreground structure. However, a shallower plateau could also result from the fact that residual foreground variance is boosted after accounting for the transfer function (see the solid red curve, left panel of Figure~\ref{fig:im_plausibility}). The onus is to argue (1) that cosmological signal power is detected and is stable to efforts to clean additional foregrounds and (2) that residual foregrounds and the signal transfer function do not conspire to appear as a stable signal band power. Additionally, there may be features in the power spectrum such as the BAO feature, or redshift-space distortions (in the 2D spectrum), that support the interpretation of cosmological signal.

Intensity mapping shares some parallels with CMB B-mode searches, where instruments must be designed to prevent mixing between bright contaminants and the signal, and foreground cleaning is a central strategy. The same language and metrics that have been developed for beam systematics in B-mode searches would be fruitfully carried over to intensity mapping. All of the lines and redshift ranges of interest have differences in experimental methodology, but the eigenvalue spectrum provides a common reference for developing instrumental requirements. Beyond astrophysical foregrounds, planning of future intensity mapping experiments should also include high-fidelity simulations of the instrument to determine requirements for the accuracy and stability of the spectral calibration.

\acknowledgements

E.S. acknowledges support as a CITA fellow, where much of this work was conducted. We thank John Ford, Anish Roshi, and the rest of the GBT staff for their support, especially in understanding instrument response. Computations were performed on the GPC supercomputer at the SciNet HPC Consortium. SciNet is funded by the Canada Foundation for Innovation.

\appendix

\section{Impact of Foreground-signal Coupling in Blind Cleaning}
\label{app:cleananticorr}

In this appendix, we find the expectation value of the correlation between the residual foregrounds in the cleaned map and the input signal. Section~\ref{ssec:pertsf} argues that the cleaned maps contain $\tilde \smap = - \mat{\Delta} \mat{S} \fgfmodes^T \fgmap - \fgfmodes \mat{S} \mat{\Delta}^T \fgmap$. The term $\mat{\Delta}$ describes how pure foreground spectral modes are influenced by the signal. In this appendix, we argue that $\tilde \smap$ is anticorrelated with the signal, on average. At first order in a perturbing signal, $\mat{\Delta} = \fgfmodes \mat{H}$ where the matrix elements are
\begin{equation}
\mat{H} |_{ij} = N_\theta^{-1} \frac{(\vect{u}^{\rm f}_i)^T (\fgmap \smap^T + \smap \fgmap^T) \vect{u}^{\rm f}_j}{\lambda^{\rm f}_j - \lambda^{\rm f}_i}.
\label{eq:pij}
\end{equation}
Here $\mat{H} = - \mat{H}^T$, is skew-symmetric because of the denominator of the perturbation element, while the numerator is symmetric by construction. The correlation between residual foregrounds in $\tilde \smap$ and signal due to spurious correlations is 
\begin{eqnarray}
\xi_{\rm spur} &=& N_\theta^{-1} \langle Tr( \tilde \smap \smap^T) \rangle = -N_\theta^{-1} \langle Tr( \fgfmodes \mat{H} \mat{S} \fgfmodes^T \fgmap \smap^T + \fgfmodes \mat{S} \mat{H}^T \fgfmodes^T  \fgmap \smap^T) \rangle \\
&=& -N_\theta^{-1} \langle Tr( \mat{H} \mat{S} \fgfmodes^T \fgmap \smap^T \fgfmodes + \mat{S} \mat{H}^T \fgfmodes^T \fgmap \smap^T \fgfmodes ) \rangle \\
&=& -N_\theta^{-1} \langle Tr(\mat{H} \mat{S} \fgfmodes^T (\fgmap \smap^T + \smap \fgmap^T)\fgfmodes ) \rangle,
\end{eqnarray}
where the first line uses $\mat{\Delta} = \fgfmodes \mat{H}$, the second line uses the cyclic property of the trace, and the third line uses the fact that a matrix and its transpose have the same trace, and the symmetry $\mat{S}^T = \mat{S}$. Let $\mat{\Sigma} = \fgfmodes^T (\fgmap \smap^T + \smap \fgmap^T)\fgfmodes$ and note that $\mat{H} |_{ij} = N_\theta^{-1} \mat{\Sigma} |_{ij} \cdot (\lambda^{\rm f}_j - \lambda^{\rm f}_i)^{-1}$. Here $\mat{\Sigma}$ is a symmetric cross-variance of the signal and foreground in the basis of the foreground modes. We can split all matrices in the foreground mode basis into cut modes and uncut modes. The matrix $\mat{S}$ is 1 along the diagonal for cut modes and zero elsewhere. Order the cuts so that they all reside in the upper left submatrix as
\begin{equation}
\mat{S} = \left( \begin{array}{cc} \mat{1}_{cc} & \mat{0}_{cu} \\ \mat{0}_{uc} & \mat{0}_{uu} \end{array} \right)~~~~~\mat{\Sigma} = \left( \begin{array}{cc} \mat{\Sigma}_{cc} & \mat{\Sigma}_{cu} \\ \mat{\Sigma}_{uc} & \mat{\Sigma}_{uu} \end{array} \right),
\end{equation}
where $c$ and $u$ denote modes that are cut vs. uncut, and analogously for $\mat{H}$. The trace evaluates to
\begin{equation}
\xi_{\rm spur} = -N_\theta^{-1} \langle Tr(\mat{H}_{cc} \mat{\Sigma}_{cc}) + Tr(\mat{H}_{uc} \mat{\Sigma}_{cu}) \rangle.
\end{equation}
The first term is the trace of the product of skew-symmetric and symmetric matrices, so it is zero. Evaluating the second term using the symmetry of $\mat{\Sigma}$ produces
\begin{equation}
\xi_{\rm spur} = -N_\theta^{-2} \left \langle \sum_{\substack{i\in cuts \\ j \notin cuts}} \frac{[(\vect{u}^{\rm f}_i)^T (\fgmap \smap^T + \smap \fgmap^T) \vect{u}^{\rm f}_j]^2}{\lambda^{\rm f}_i - \lambda^{\rm f}_j} \right \rangle, 
\end{equation}
taking the transpose of $\mat{H}$ and reversing the denominator to preserve sign. The elements of the spurious correlation $\mat{\Sigma} |_{ij}$ have zero mean (over signal realizations), but the correlation between signal and cleaning residuals in $\xi_{\rm spur}$ appears quadratically inside the average over signal realizations, resulting in a net bias.

Take the SVD of $\fgmap$, $\fgmap = \sqrt{N_\theta} \fgfmodes \mat{\Lambda}_{\rm f}^{1/2} \fgsmodes^T$ so that
\begin{equation}
\xi_{\rm spur} = -N_\theta^{-1} \left \langle \sum_{\substack{i\in cuts \\ j \notin cuts}} \frac{[\sqrt{\lambda^{\rm f}_i} (\vect{v}^{\rm f}_i)^T \smap^T \vect{u}^{\rm f}_j + \sqrt{\lambda^{\rm f}_j} (\vect{u}^{\rm f}_i)^T \smap \vect{v}^{\rm f}_j]^2}{\lambda^{\rm f}_i - \lambda^{\rm f}_j} \right \rangle 
\end{equation}
By the construction of the filter, $\lambda^{\rm f}_i \gg \lambda^{\rm f}_j$ for $i \in {\rm cuts}$ and $j \notin {\rm cuts}$ because the cuts remove the highest variance foreground modes. In this limit, 
\begin{equation}
\xi_{\rm spur} = -N_\theta^{-1} \left \langle \sum_{\substack{i\in cuts \\ j \notin cuts}} [ (\vect{v}^{\rm f}_i)^T \smap^T \vect{u}^{\rm f}_j]^2 \right \rangle.
\end{equation}
The amplitudes of foregrounds drop out, and the matrix element $(\vect{v}^{\rm f}_i)^T \smap^T \vect{u}^{\rm f}_j$ is the overlap of the signal with the subtracted foreground spatial modes and unsubtracted spectral modes. If the subtracted modes have a smooth spatial distribution, spurious correlations will wipe out signal at low $k_\perp$. 
Using the SVD of the signal $\smap = \sum_n \sqrt{\lambda^{\rm s}_n N_\theta} \vect{u}^s_i (\vect{v}^s_i)^T$,
\begin{equation}
\xi_{\rm spur} = -\left \langle \sum_{\substack{i\in cuts \\ j \notin cuts}} \left [ \sum_n \sqrt{\lambda^{\rm s}_n}(\vect{v}^{\rm f}_i)^T \vect{v}^s_n  (\vect{u}^s_n)^T \vect{u}^{\rm f}_j \right ]^2 \right \rangle. 
\end{equation}
To expand the squared sum on $n$, note that cross terms with $n \neq n'$ will average to zero in the $\langle \rangle$ over signal. Recall that both the spatial and spectral modes are normalized so that $\vect{u}^T \vect{u} = 1$ and $\vect{v}^T \vect{v} = 1$. To get a rule of thumb, let $N_{res, \nu}$ be the effective number of spectral degrees of freedom of the signal fluctuation. Then each inner product $[(\vect{u}^s_k)^T \vect{u}^{\rm f}_j]^2 \approx 1/N_{res, \nu}$ and the spatial inner products-squared scale as $\approx 1/N_{res, \theta}$. Let the sum on $i \in {\rm cuts}$ be over $N_m$ cut modes. If there are $N_{res, \nu}$ spectral modes available in the survey, then the number of uncut modes in the sum $j \notin cuts$ is over $N_{res, \nu} - N_m$. Using the sum of the signal eigenvectors from Equation~\ref{eqn:sigcov}, the scaling of $\xi_{\rm spur}$ gives the rule of thumb
\begin{equation}
\xi_{\rm spur} \approx -\xi_{\rm s} \frac{N_m (N_{res, \nu} - N_m)}{N_{res, \nu} N_{res, \theta}}.
\end{equation}
Recall that $\xi_{\rm clean} = \xi_{\rm s} + \xi_{\rm direct} + \xi_{\rm spur}$, or
\begin{equation}
\xi_{\rm clean} = \xi_{\rm s} \left [1 - \frac{N_m}{N_{res, \nu}} - \frac{N_m (N_{res, \nu} - N_m)}{N_{res, \nu} N_{res, \theta}} \right ] = \left ( 1 - \frac{N_m}{N_{\rm res, \nu}} \right ) \left ( 1-\frac{N_m}{N_{\rm res, \theta}} \right ) \xi_s.
\end{equation}
In interpreting this rule of thumb, it is useful to think about the survey volume filtered onto particular scales for a given band power. The terms $N_{\rm res, \nu}$ and $N_{\rm res, \theta}$ are essentially unrelated to the number of frequency bins $N_\nu$ and spatial pixels $N_\theta$ in the survey. Either $N_\nu$ or $N_\theta$ could be made arbitrarily large through mapping with finer pixels or a larger number of spectral channels. Instead, the relevant quantity is the number of spatial and spectral degrees of freedom that the signal in the given 2D band power can explore. If there can be many signal realizations on a given scale, then the spurious correlation with the foregrounds averages down better.

\section{Relation of the Cross-power to the Optimal Estimator}
\label{app:crosspower}

Split the season into two maps $\vect{x}^T = (\vect{x}_A, \vect{x}_B)^T$ and let the covariance be
\begin{equation}
\mat{C} = \left ( \begin{array}{cc} \mat{S} + \mat{N}_A & \mat{S}_\times \\  \mat{S}_\times & \mat{S} + \mat{N}_B \end{array} \right ), 
\end{equation} 
where $\mat{S} = \mat{S}_\times = \sum_\alpha p_\alpha \mat{S}_{,\alpha}$ is the signal covariance and $\mat{N}_A$, $\mat{N}_B$ are the noise covariance in the two maps. We assume that the noise covariance contains only thermal noise, which is uncorrelated between subseasons. Any signal on the sky (including residual foregrounds) is correlated between subseasons, and we absorb it in the signal covariance. The optimal estimator remains
\begin{equation}
\hat q_\alpha \propto \vect{x}^T \mat{C}^{-1} \mat{C}_{,\alpha} \mat{C}^{-1} \vect{x}.
\end{equation}

This generically involves combinations of the data like $\vect{x}_A^T \mat{Q} \vect{x}_A$ (auto-powers) and $\vect{x}_A^T \mat{Q} \vect{x}_B$ (cross-powers). To avoid noise bias, we would like to avoid terms like $\vect{x}_A^T \mat{Q} \vect{x}_A$. This is done by 1) neglecting the blocks along the diagonal of $\mat{C}_{,\alpha}$, and 2) by neglecting the signal covariance contribution to $\mat{C}$. These choices are
\begin{equation}
\mat{C}_{,\alpha}\biggl |_{cross} = \left ( \begin{array}{cc} 0 & \mat{S}_{,\alpha} \\  \mat{S}_{,\alpha} & 0 \end{array} \right ),~~~~\mat{C}\biggl |_{cross} = \left ( \begin{array}{cc} \mat{N}_A & 0 \\  0 & \mat{N}_B \end{array} \right ) \Rightarrow b_\alpha = Tr(\mat{C}_{,\alpha} \mat{C}) = 0.
\end{equation}
Putting these factors together, the crossed estimator is 
\begin{equation}
\hat q_\alpha \propto (\mat{N}_A^{-1} \vect{x}_A)^T \mat{S}_{,\alpha} (\mat{N}_B^{-1} \vect{x}_B)^T.
\end{equation}

Formally, this cross-power is suboptimal because it neglects signal correlations in the weighting, and it neglects signal information in the auto-power. 

\section{Gaussian Errors}
\label{app:gauss_error}

This derivation follows \citet{2011ApJ...729...62D} except that we do not form explicit map differences to estimate thermal noise. The covariance of an estimator $\hat P_{i\times j}$ across Gaussian fields $i, j$ is 
\begin{equation}
{\rm cov} (\hat P_{i\times j}, \hat P_{k\times l}) = \frac{1}{\nu(k)} \left [ \langle P_{i\times k} \rangle \langle P_{j\times l} \rangle + \langle P_{i\times l} \rangle \langle P_{k\times j} \rangle \right ], 
\label{eqn:generalcov}
\end{equation}
where $\nu(k)$ is the effective number of modes that enter the average for the band power. Model the power spectra as $\langle P_{i\times j} \rangle = P_{\rm auto}$ for $i=j$ and $\langle P_{i\times j} \rangle = P_\times$ for $i \neq j$. For simplicity, we will assume that each map section has approximately the same statistical properties so that the cross-powers are represented by $P_\times$ (e.g. $A \times B$) and the auto-powers are represented by $P_{\rm auto}$ (e.g. $A \times A$) to a good approximation. However, in surveys where map sections have different integration times or noise properties, these expressions should be expanded to break out the noise properties of the different subsurveys. Note that $P_{\rm auto}$ includes both thermal noise and sky variance, and $P_\times$ includes any sky variance (including residual foregrounds).

The covariance in Equation\ref{eqn:generalcov} of several data combinations is
\begin{eqnarray}
{\rm no~sec.~in~common} & \frac{2}{\nu(k)} P_\times^2 \\
{\rm one~sec.~in~common} & \frac{1}{\nu(k)} \left [ P_\times^2 + P_\times P_{\rm auto} \right ] \\
{\rm two~sec.~in~common} & \frac{1}{\nu(k)} \left [ P_\times^2 + P_{\rm auto}^2 \right ].
\end{eqnarray}
An example of the first case would be $AB, CD$, the second case would be $AB, AC$ and the third case, $AB, AB$. The total covariance of the estimated power spectrum is the sum of the covariance cases above with appropriate multiplicities 
\begin{eqnarray}
{\rm cov} (\hat P(k), \hat P(k)) &=& \frac{1}{N_s (N_s-1)} \left [ (N_s -2)(N_s - 3) \frac{2}{\nu(k)} P_\times^2 + 4 (N_s - 2) \frac{1}{\nu(k)} ( P_\times^2 + P_\times P_{\rm auto} )  + \frac{2}{\nu(k)} ( P_\times^2 + P_{\rm auto}^2 ) \right ] \nonumber \\
&=& \frac{1}{\nu(k) N_s (N_s-1)} \left [  2 (N_s^2 - 3 N_s +3) P_\times^2 + 4 (N_s -2) P_\times P_{\rm auto} + 2 P_{\rm auto}^2\right ]. 
\label{eqn:finalgauss}
\end{eqnarray}
We can cast this in a more familiar form by letting $P_\times = P_{\rm sig}$ be the ``signal" and $P_{\rm auto} = P_{\rm sig} + P_{\rm n}$ ``signal plus noise." Separating the signal and noise powers from $P_{\rm auto}$ one has
\begin{equation}
\nu(k) {\rm cov} (\hat P(k), \hat P(k)) = 2 P_{\rm sig}^2 + 4 \frac{P_{\rm sig} P_n}{N_s} + 2 \frac{P_{\rm n}^2}{N_s (N_s-1)}.
\end{equation} 

These Gaussian errors determine the diagonal of the bandpower covariance and require an estimate of the number of modes $\nu(k)$. In practice, restrictions of the survey volume due to spatial/spectral weighting and masking produce a complex covariance structure and effective number of modes. We advocate a hybrid approach where the bandpower covariance diagonal is estimated through Gaussian errors using the auto- and cross-powers of the data (and so represents sample variance and thermal noise). This is then used to calibrate a full covariance matrix determined by Monte Carlo of the complete pipeline.

Let $\mat{C}_{\rm sim} = {\rm cov} (\hat P(k)_{\rm sim}, \hat P(k')_{\rm sim})$ be the measured covariance of signal plus noise simulations of the data pipeline. Put the square root of Gaussian errors derived for the simulation (Equation~\ref{eqn:finalgauss}) along the diagonal of $\mat{\Lambda}_{\rm sim}$ and likewise for the measured data $\mat{\Lambda}_{\rm data}$. Then, recalibrate the band power covariance measured in simulations against the variance measured in the data as
\begin{equation}
\mat{C}_{\rm data} = \mat{\Lambda}_{\rm data} \mat{\Lambda}_{\rm sim}^{-1} \mat{C}_{\rm sim} \mat{\Lambda}_{\rm sim}^{-1} \mat{\Lambda}_{\rm data}.
\end{equation}
In the operation $\mat{\Lambda}_{\rm sim}^{-1} \mat{\Lambda}_{\rm data}$, the common factor $\nu(k)$ drops out of the Gaussian errors of the data and simulations and does not need to be calculated explicitly. The simulation pipeline does not need a high-fidelity model of the real data's covariance. Instead, the goal of the simulations is to measure the off-diagonal terms, and the $A\times A$ combinations of the real data give Gaussian errors that calibrate the amplitudes.

\bibliographystyle{apj}
\bibliography{intensity_methods}

\end{document}